\documentclass[%
 reprint,
 amsmath,amssymb,
 aps,longbibliography,
]{revtex4-1}

\usepackage{graphicx}
\usepackage{subfig}
\usepackage{color}
\usepackage{dcolumn}
\usepackage{bm}

\newcommand{\Par}{\partial}
\newcommand{\vare}{\varepsilon}

\newcommand{\abs}[1]{\left| #1 \right|}
\newcommand{\set}[1]{\left\{#1 \right\}}

\newcommand{\INF}{\infty}

\newcommand{\RE}{\mathrm{Re}}
\newcommand{\IM}{\mathrm{Im}}
\newcommand{\PT}{\mathcal{PT}}
\begin{document}

\preprint{APS/123-QED}

\title{Exceptional Points of Resonant States on a Periodic Slab}

\author{Amgad Abdrabou}
 \affiliation{Department of Mathematics, City University of Hong Kong, Kowloon, Hong Kong, China}
\author{Ya Yan Lu}%
 \email{Corresponding author: mayylu@cityu.edu.hk}
\affiliation{Department of Mathematics, City University of Hong Kong, Kowloon, Hong Kong, China}%

\date{\today}

\begin{abstract}
A special kind of degeneracies known as the exceptional points (EPs),
for resonant states on a dielectric periodic slab, are investigated.
Due to their unique properties, EPs have found important applications
in lasing, sensing, unidirectional operations, etc. In general, EPs
may appear in non-Hermitian eigenvalue problems, including those
related to $\mathcal{PT}$-symmetric systems and those for open
dielectric structures (due to the existence of radiation loss). In
this paper, we study EPs on a simple periodic structure: a slab with a
periodic array of gaps. Using an efficient numerical method, we
calculate the EPs and study their dependence on geometric parameters.
Analytic results are obtained for the limit as the periodic slab
approaches a uniform one.  Our work provides a simple platform for
further studies concerning EPs on dielectric periodic structures, their
unusual properties and applications. 
\begin{description}
\item[PACS numbers]
42.65.Hw,42.25.Fx,42.79.Dj
\end{description}
\end{abstract}

\maketitle


\section{Introduction}\label{sec1}

In parameter-dependent eigenvalue problems of non-Hermitian operators,
a special kind of degeneracy may occur at some particular values of
the system parameters, that is, two or more eigenvalues coalesce and
their corresponding eigenfunctions collapse into one single function.
Such a spectral degeneracy is called an exceptional point
(EP)~\cite{Kato}. 
The EPs are interesting, because they give rise to unusual
physical phenomena in many systems related to 
non-Hermitian eigenvalue problems
\cite{Moiseyev,Heiss,Jones,HeissChaos1,HeissChaos2,Leyvarz,Holger}. 
The EPs are also central to quantum or classical parity-time
($\mathcal{PT}$) symmetric systems~\cite{BenderPT,BenderPT2}.
In these systems, an EP corresponds to a transition (also known as the $\mathcal{PT}$
symmetry breaking) from a state with all real eigenvalues to a 
state with complex eigenvalues~\cite{BenderPT3}. 
In recent years, $\mathcal{PT}$-symmetric optical systems have been
intensively investigated. EPs have been observed in
 $\PT$-symmetric waveguides~\cite{Klaiman,PTsym}, and 
 exploited in a variety of optical systems leading to a number of unusual wave
 phenomena and novel applications such as the revival of
 lasing~\cite{EPLas1,EPLas2,EPLas3}, enhanced 
 sensing~\cite{Hodaei,Chen2017}, stopping light pulses~\cite{Goldzak},
 single mode lasers~\cite{Feng}, and unidirectional
 invisibility~\cite{Lin}. 

Due to the outgoing radiation conditions, the eigenvalue problem of
resonant modes on open structures is non-Hermitian, even
if the dielectric function $\vare$ of the structure is real and
positive. Therefore, EPs of resonant modes could exist on properly
designed passive structures. For example, a micro-toroid cavity with
two nearby nanoscale scatterers was designed to have an EP, and
it was used to enhance sensing \cite{LYang}. 
Periodic structures surrounded by or sandwiched between free space are
also open structures. Thus, the eigenvalue problem for resonant
modes on such a periodic structure is non-Hermitian and there could be
EPs. In fact,  EPs have been observed on a photonic crystal slab
\cite{BoZhen,Arslan}, and they exist for any specified wavevector
direction, as far as the geometric parameters are properly chosen. 

To reveal novel properties of EPs on open dielectric structures and
realize their potential applications, it is necessary to carry out
systematic studies on EPs. In this paper, we develop an efficient
numerical method for computing second-order (i.e., doubly degenerate)
EPs based on the square-root splitting of the eigenvalues, calculate
the EPs on a simple dielectric periodic slab, and find out how they
vary with the geometric parameters.  Although the structure is very
simple, many EPs exist, and they exhibit rather complicated dependence
on the parameters. However, we are able to find a simple analytic
result for the limit as the periodic slab approaches a uniform one.
The EPs in this limit are related to artificial degeneracies of the
guided modes of the uniform slab when it is regarded as a periodic
one.  The rest of this paper is organized as follows. In
Sec.~\ref{sec2}, we give some definitions and show the band structure
of a particular periodic slab. In Sec.~\ref{S3}, we develop an
efficient numerical method for computing EPs, and show the band
structure of a periodic slab with one EP. In Sec.~\ref{S4}, we analyze
the artificial degeneracies of a uniform slab. In Sec.~\ref{S5}, we
calculate the EPs on the periodic slab and show their dependence on
parameters, including the limit studied in Sec.~\ref{S4}. Finally, we
conclude our paper with some remarks in Sec.~\ref{S6}. 

\section{Resonant modes}\label{sec2}

\begin{figure}[htb]
	\centering 
	\includegraphics[width=0.90\linewidth]{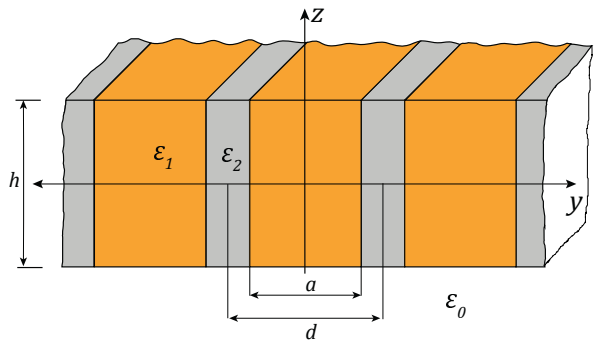}
	\caption{A dielectric slab which is invariant in $x$ and periodic
          in $y$. Each period consists of two segments.}
	\label{Fig1}
\end{figure}

A typical periodic dielectric slab is shown in 
Fig.~\ref{Fig1}. 
The structure is invariant in $x$, periodic in $y$ with period $d$, finite in
 $z$ with a thickness $h$, and surrounded by air. It is further
 assumed that each period of the slab consists of two segments with 
 dielectric constants 
 $\vare_1$ and $\vare_2$, and widths $a$ and $d-a$, respectively. 
%
For the $E$-polarization, the $x$ component of the electric field,
denoted as $u$, satisfies the following two dimensional (2D) Helmholtz
equation:  
\begin{equation}\label{Eq1}
\partial_y^2 u +\partial_z^2 u +k^2 \vare(y,z)u = 0,
\end{equation}
where $\vare=\vare(y,z)$ is the dielectric function of the structure,
$k=\omega/c$ is the free-space wavenumber, $\omega$ is the angular
frequency, and $c$ is the speed of light in vacuum. 

A Bloch mode on the periodic slab is a solution of Eq.~\eqref{Eq1}
given in the form 
\begin{equation}\label{Eq2}
u(y,z) = \phi(y,z)\,e^{i\beta y},
\end{equation}
where $\beta$ is a real Bloch wavenumber satisfying $\abs{\beta}\leq
\pi/d$, and $\phi(y,z)$ is periodic  in $y$ with period
$d$. In the free space surrounding the slab, i.e., for $\abs{z}> h/2$,
the solution can  be expanded in plane waves as 
\begin{equation}\label{Eq3}
u(y,z) = \sum_{m=-\INF}^{\INF} \hat{u}_{m}^{\pm} e^{i(\beta_m y\pm\alpha_m z)}, \quad \pm z > h/2,
\end{equation}
where $\hat{u}_m^\pm$ are the expansion coefficients, 
\begin{equation}\label{Eq4}
\beta_m = \beta+2\pi m/d,\quad \alpha_m  =\sqrt{k^2-\beta_m^2},
\end{equation}
 and the square root is defined with a branch cut along the
 negative imaginary axis.

If $\phi(y,z) \to 0$  as $\abs{z} \to \INF$,
then the Bloch mode is a guided mode. 
Below the light line, i.e., for $ k < \abs{\beta}$,
guided modes exist continuously with respect to the frequency and the
wavenumber. Above the light line, Bloch modes with the expansion
(\ref{Eq3}) are typically resonant modes 
with a complex frequency, that is, $k$ is complex and
$\RE(k)>\abs{\beta}$.  The resonant modes satisfy 
outgoing radiation conditions as $\abs{z} \to \INF$. From
Eq.~\eqref{Eq4}, it is clear that $\alpha_0$ is a complex
number with a negative imaginary part. Therefore, the plane waves
$\mathrm{exp}[i(\beta y\pm \alpha_0 z)]$ blow up as $ z \to \pm \INF$,
respectively.  The quality factor, denoted by $Q$, of a
resonant mode 
is given by $Q = -0.5\, \mathrm{Re}(k)/\mathrm{Im}(k)$. 
In special circumstances, resonant modes with infinite quality factors, i.e. $\mathrm{Im}(k) = 0$, may exist, and they are the bound
states in the continuum (BICs). The BICs 
have intriguing properties and important applications
\cite{BICreview}.
 
Numerical methods for computing the Bloch modes can be classified as
linear and nonlinear schemes. A linear scheme discretizes the Helmholtz
equation directly to obtain a linear matrix eigenvalue problem for
eigenvalue $k^2$. A nonlinear scheme produces a nonlinear 
eigenvalue problem with a smaller matrix whose entries depend on 
$k$ implicitly. The mode matching method is a nonlinear scheme. 
For the structure shown in Fig.~\ref{Fig1}, 
it gives rise to a homogeneous linear system 
 ${\bm A}(\beta,k) {\bm x}={\bm 0}$, where ${\bm x}$ is a vector
 of unknown expansion coefficients.
 For a given $\beta$, nontrivial
 solutions can be found by searching complex $k$ such that
 $\lambda_1({\bm A}) = 0$, where $\lambda_1({\bm A})$ is the
 eigenvalue of ${\bm A}$ with the smallest magnitude. Additional
 details about the numerical methods are given in Appendix A. 
 
\begin{figure}[htb]
 	\centering
 	\includegraphics[width=0.8\linewidth]{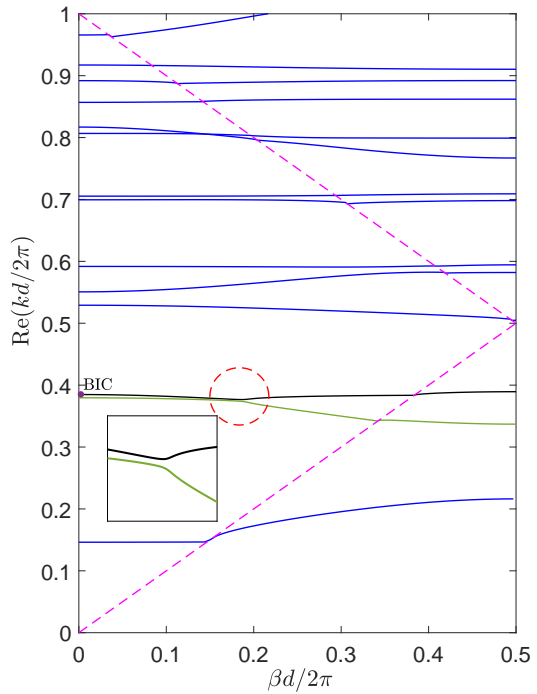}
 	\caption{Band structure of Bloch modes (odd in $z$) on a
          periodic slab with $\vare_1 = 15.42$, $\vare_2 = 1$, $a =
          0.5d$ and $h = 1.16d$.}\label{Fig4} 
 \end{figure}

\begin{figure}[htbp]
 	\centering 
 	\includegraphics[width=0.8\linewidth]{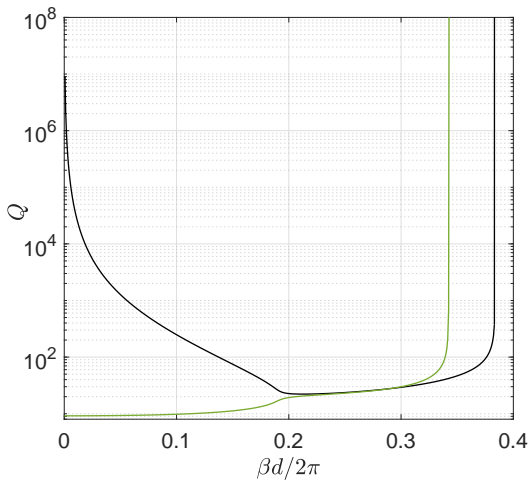}
 	\caption{Quality factors of the resonant modes corresponding
          to the black and green curves in Fig.~\ref{Fig4}.}
\label{Fig5}
 \end{figure}

As a numerical example, we show the band structure of a periodic slab
with $\vare_1 = 15.42$,   $\vare_2 = 1$,  $a = 0.5d$ and $h = 1.16d$
in Fig.~\ref{Fig4}.  
To obtain these results, we first use a linear scheme to calculate the
eigenmodes at $\beta=0$, and then use the more accurate mode matching 
method to find each band for $0<\beta\leq \pi/d$. For simplicity, we
only show $\RE(k)$ for resonant modes that are odd functions of $z$. 
Notice that two curves, the solid black one and the solid green one,
are close to each other at $\beta \approx 0.18(2\pi/d)$. 
  In Fig.~\ref{Fig5}, 
we show the quality factors of the resonant modes corresponding to 
these two curves. It can be seen that the quality factors are also
close to each other at $\beta\approx 0.18(2\pi/d)$. Notice that the quality
factor of the black curve diverges at $\beta= 0$, and this corresponds
to a BIC at $\beta=0$, i.e., a standing wave. Meanwhile, the quality factors of
both curves diverge as the pair $(\beta, k)$ approaches the light line. 

In the next section, we show that by tuning one structural 
  parameter, these two resonant modes can be forced to coalesce,
  giving rise to a second-order EP.

\section{Exceptional Points\label{S3}}

If two eigenvalues, say $k_+$ and $k_-$, are close to each other for
some $\beta$, there may be an EP on a structure with slightly
different parameters. To find the EP, we can try to find the parameter
values and $\beta$, such that $k_+ = k_-$, and check whether the
eigenfunctions also coalesce.  This approach is tedious and
inefficient. In the following, we develop an efficient method based on
the square-root splitting of the eigenvalues at second-order
EPs. 


Let $k_*$ and $\beta_*$ be the eigenvalue and Bloch wavenumber of a
second-order EP associated with the bands
$k_+(\beta)$ and $k_-(\beta)$. In the vicinity of the EP, 
the eigenvalues have the following approximations
\begin{eqnarray}
\label{Eq5}
&&    k_{\pm}(\beta) \approx k_*\pm (b_1+i\,b_2)\sqrt{\beta-\beta_*},
  \quad \beta>\beta_*,  \\
\label{Eq6}
&&        k_{\pm}(\beta) \approx k_*\pm
   (c_1+i\,c_2)\sqrt{\beta_*-\beta},  \quad \beta < \beta_*,
\end{eqnarray}
where $b_1$, $b_2$, $c_1$ and $c_2$ are unknown real constants.
Since the EP is a resonant mode with $\beta =\beta_*$ and
$k = k_*$, we have 
 \begin{equation}\label{Eq15a}
\lambda_1\left[{\bm A}(\beta_*,k_*) \right] = 0,
 \end{equation}
where ${\bm A}(\beta,k)$ is a matrix obtained by the mode matching
method (see Appendix A). In addition, for a small $\delta\beta>0$, two
resonant modes exist at $\beta = \beta_*+\delta\beta$ with
$k_{\pm}(\beta)$  given by Eq.~\eqref{Eq5}. This leads to 
\begin{eqnarray}
\label{Eq15b}
&& \lambda_1\left[ {\bm A}(\beta_*+\delta\beta,k_*+\delta k )
   \right] \approx 0,  \\
&& \label{Eq15c}
  \lambda_1\left[{\bm A}(\beta_*+\delta\beta,k_*-\delta k) \right]
   \approx 0,   
\end{eqnarray}
where $\delta k = (b_1+ib_2)\sqrt{\delta\beta}$. 
Therefore, we can find second-order EPs by choosing a proper
$\delta\beta$ and solving
Eqs.~\eqref{Eq15a}, \eqref{Eq15b} and \eqref{Eq15c}. 
It turns out that EPs can be found by tuning just one structural
parameter. If $\vare_1$, $\vare_2$ and $a$ are fixed, we can search the
slab thickness $h$ to find EPs. The three complex equations 
\eqref{Eq15a}, \eqref{Eq15b} and \eqref{Eq15c}, corresponding to six real
equations, are used to determine six real unknowns: 
$\set{h_*, \beta_*, \RE(k_*), \IM(k_*), b_1, b_2}$, where $h_*$
denotes the particular value of $h$ for EPs. The values of $b_1$ and
$b_2$ describe how strongly the modes split from the EP as
$\beta$ moves away from $\beta_*$. 

Following the example in Sec.~\ref{sec2}, 
we choose $\vare_1=15.42$, $\vare_2 = 1$ and $a=0.5d$, and allow $h$
to vary. Using the method described above, we found an EP 
 for  $h_*=1.154485\,d$, $\beta_* = 0.187005(2\pi/d)$ and
$k_*=(0.3761904-0.0092194i)\,(2\pi/d)$. 
The band structure of the period slab with thickness $h_*$ 
is shown in 
\begin{figure}[htb]
  \centering 
	\includegraphics[width=0.8\linewidth]{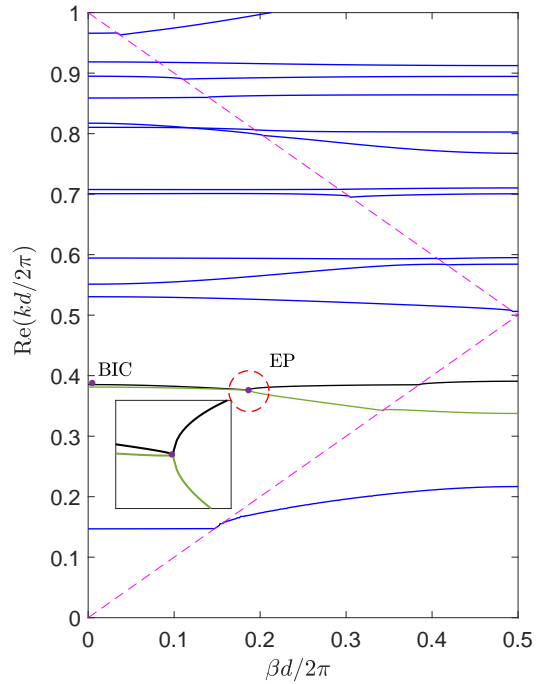}
	\caption{Band structure of Bloch modes on a periodic slab with
          $\vare_1 = 15.42$, $\vare_2 = 1$, $a = 0.5d$ and $h_*= 1.154485d$.}
\label{Fig6}	
\end{figure}
Fig.~\ref{Fig6}. It is clear that two two bands touch at the EP with a square-root 
splitting in the vicinity of $\beta_*$. The quality factors of these two 
bands are shown in Fig.~\ref{Fig7}. 
   \begin{figure}[htb]
	\centering
	\includegraphics[width=0.8\linewidth]{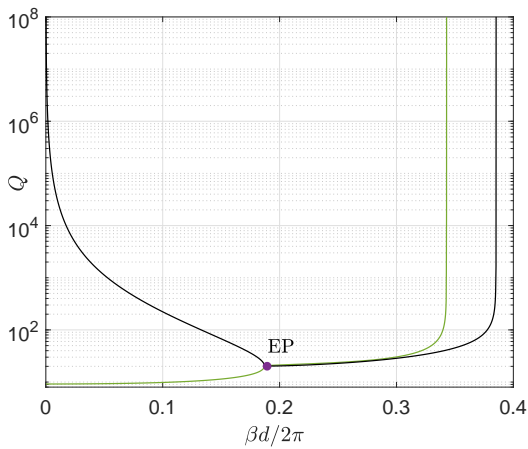}
	\caption{Quality factors of the resonant modes corresponding
          to black and green curves in Fig.~\ref{Fig6}.}\label{Fig7}
\end{figure}
As expected, the two curves in Fig.~\ref{Fig7} also touch at $\beta_*$.

 It should be pointed out that an EP can be calculated by using either 
 Eq.~\eqref{Eq5} or Eq.~\eqref{Eq6}. For the latter case, we choose a
 negative $\delta \beta$ and calculate the constants $c_1$ and $c_2$. It
 turns out that these constants  satisfy 
\begin{equation}\label{EqSW}
c_1 = \pm b_2 ,\quad\mathrm{and}\quad  c_2 = \mp b_1. 
\end{equation}

\begin{figure*}[htb]
	\subfloat{\includegraphics[width=0.33\textwidth]{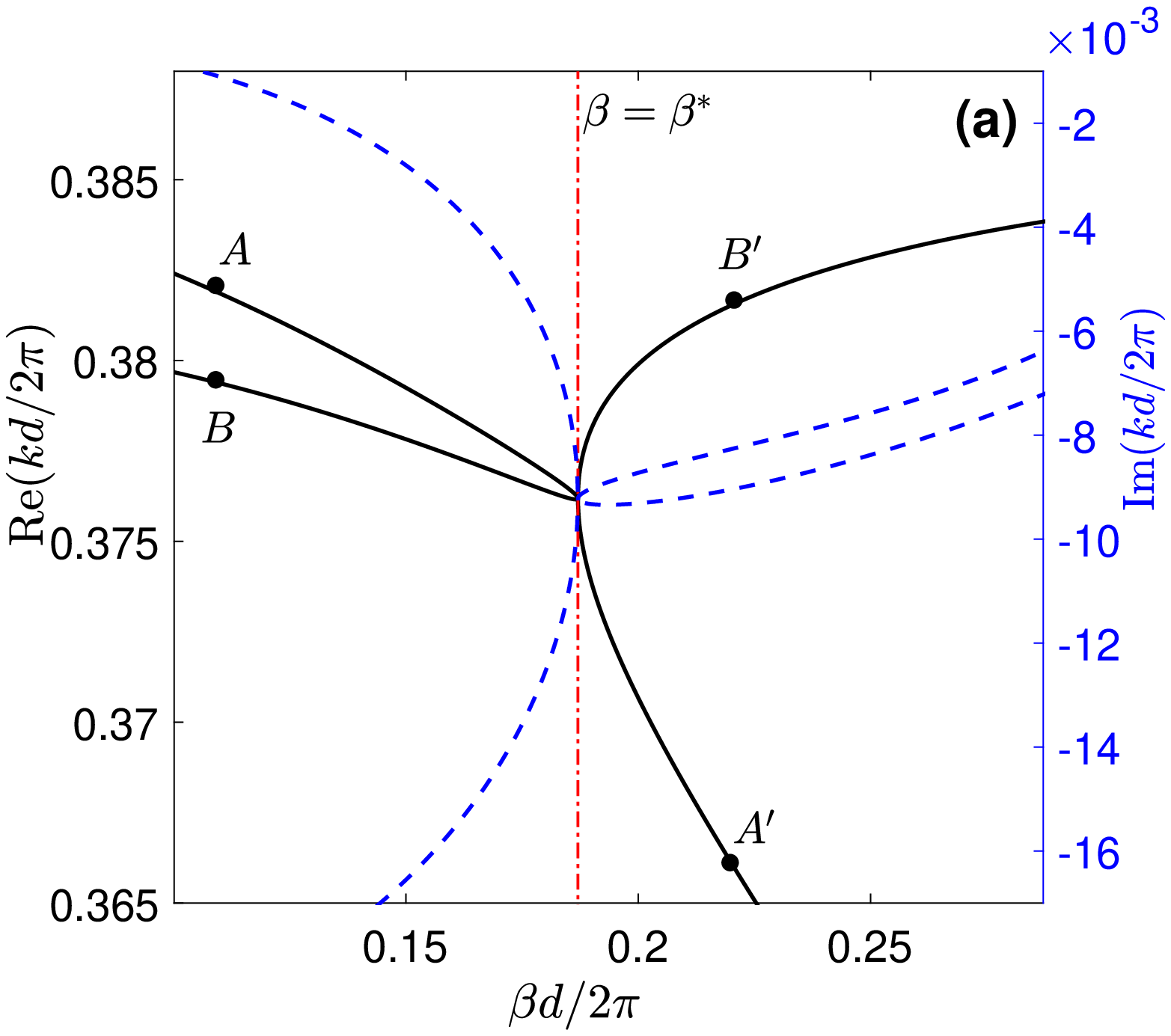}}\hfill
	\subfloat{\includegraphics[width=0.33\textwidth]{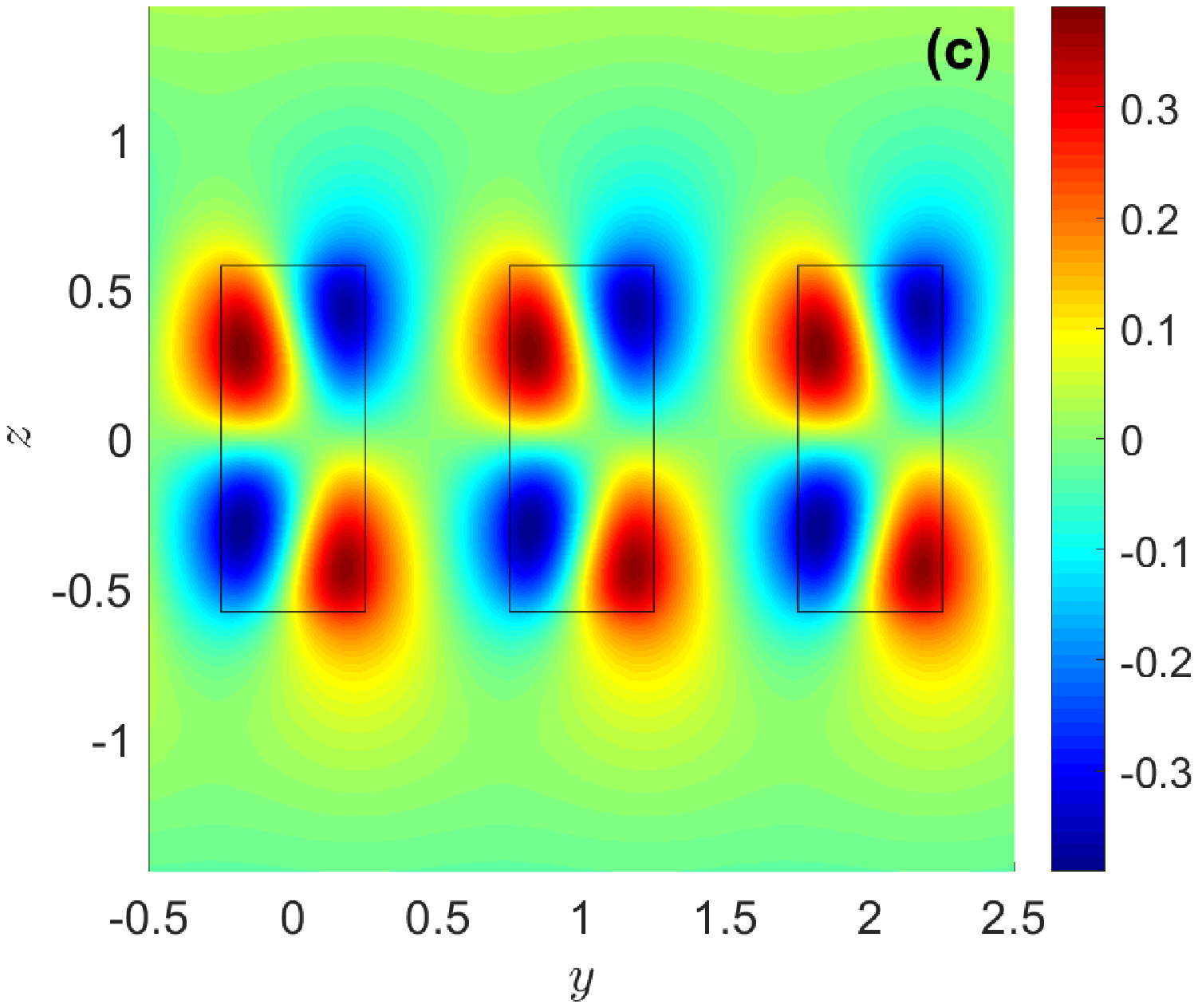}}\hfill
	\subfloat{\includegraphics[width=0.33\textwidth]{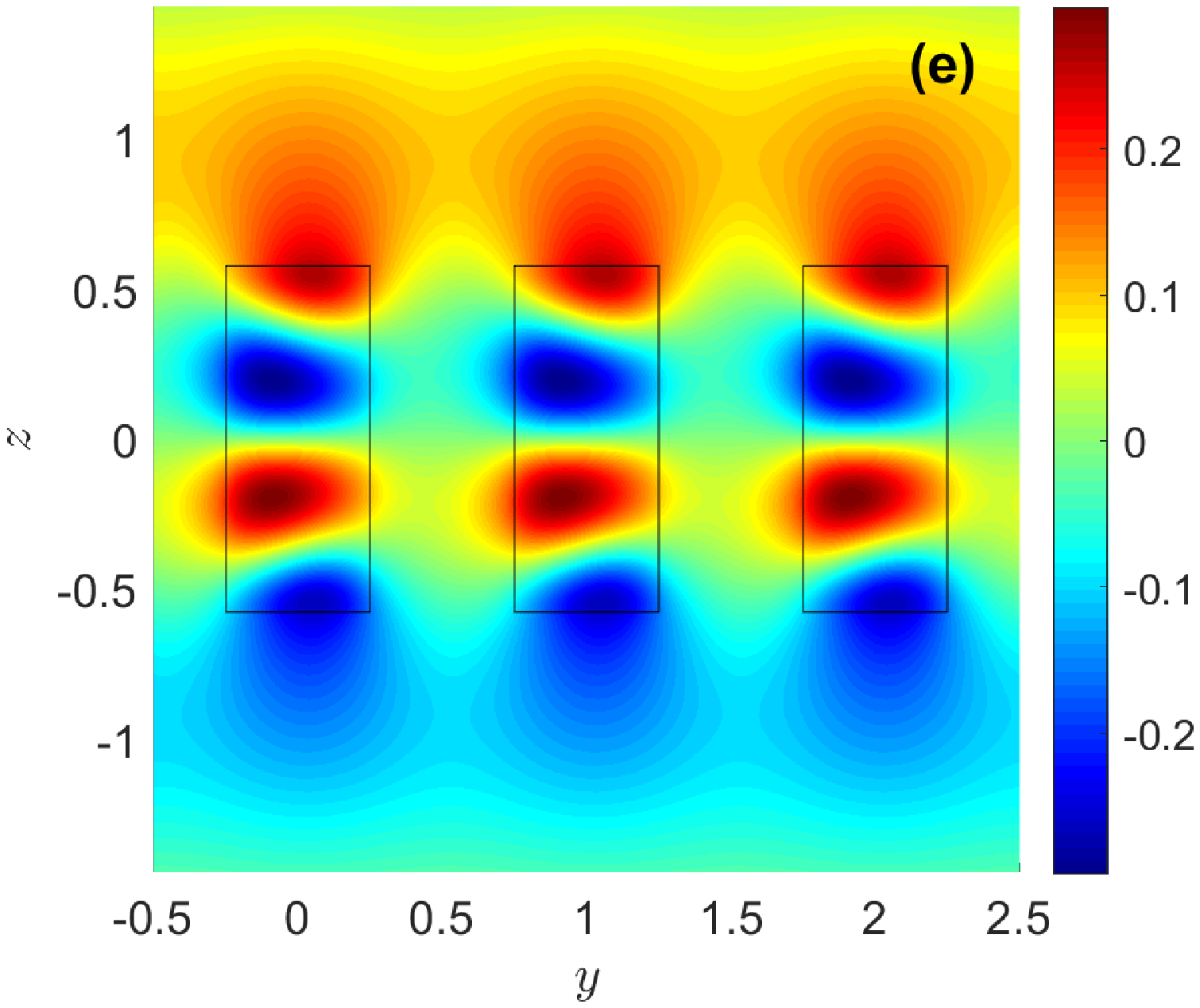}}\\[-2ex]  
	\subfloat{\includegraphics[width=0.33\textwidth]{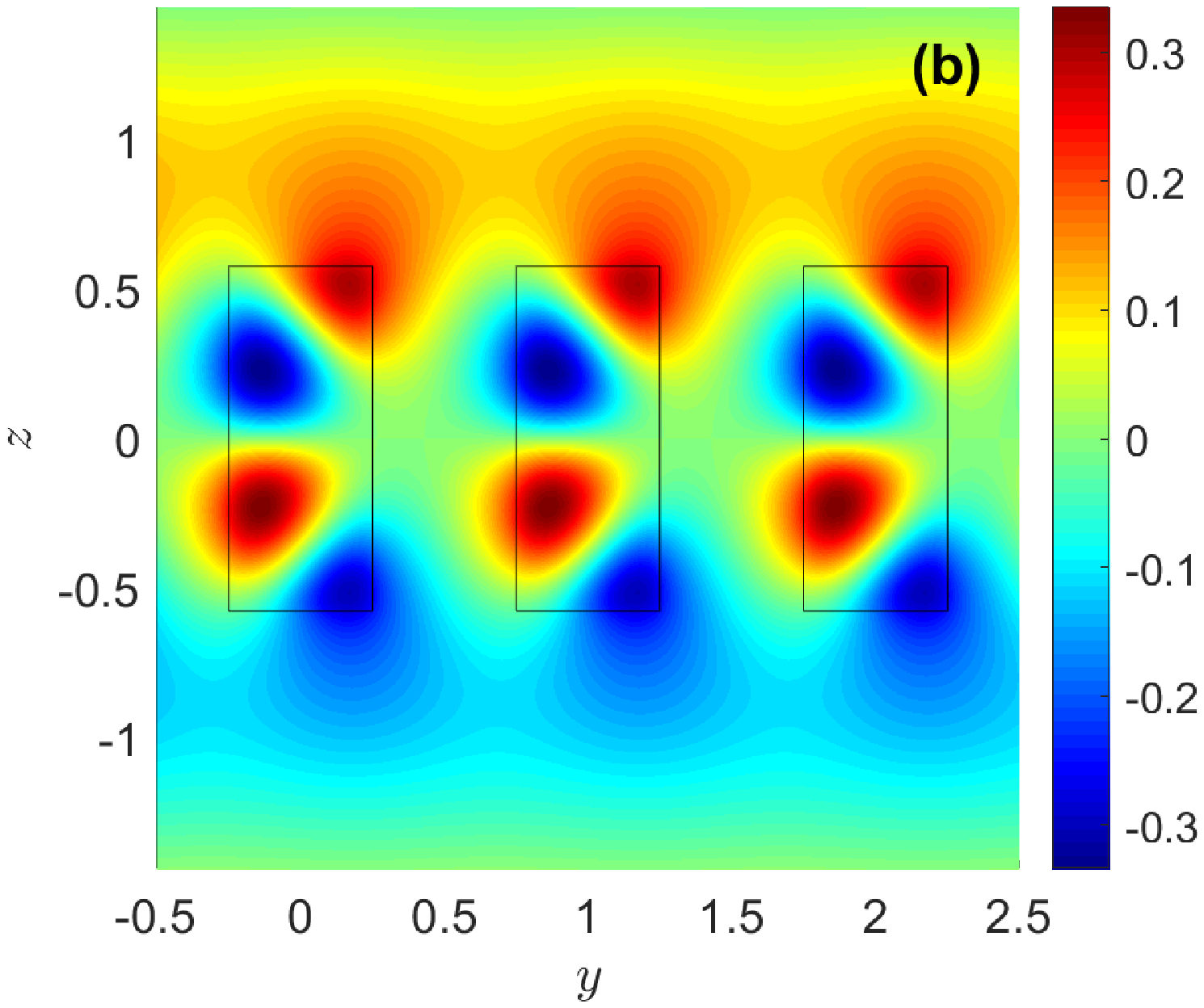}}\hfill
	\subfloat{\includegraphics[width=0.33\textwidth]{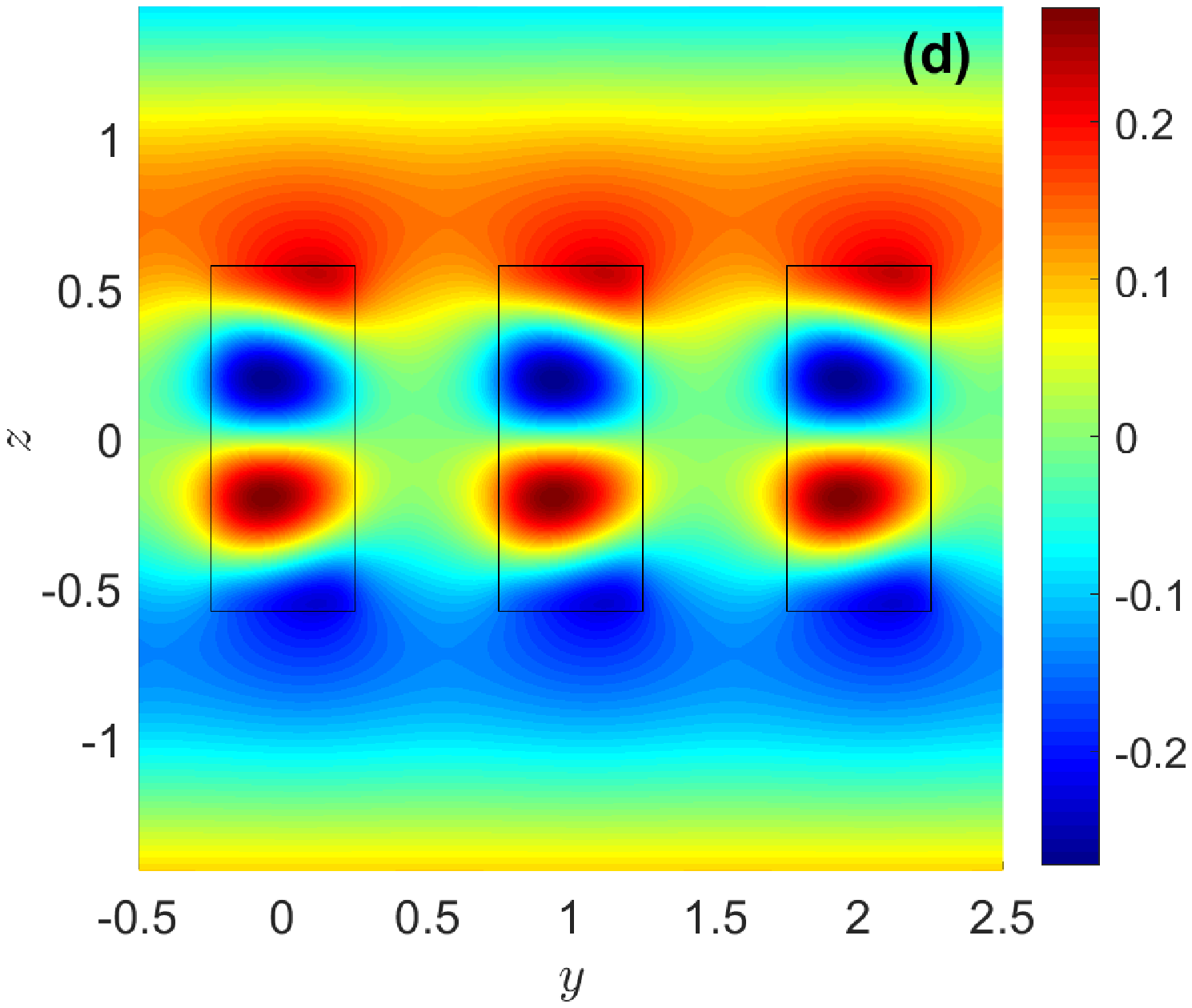}}\hfill
	\subfloat{\includegraphics[width=0.33\textwidth]{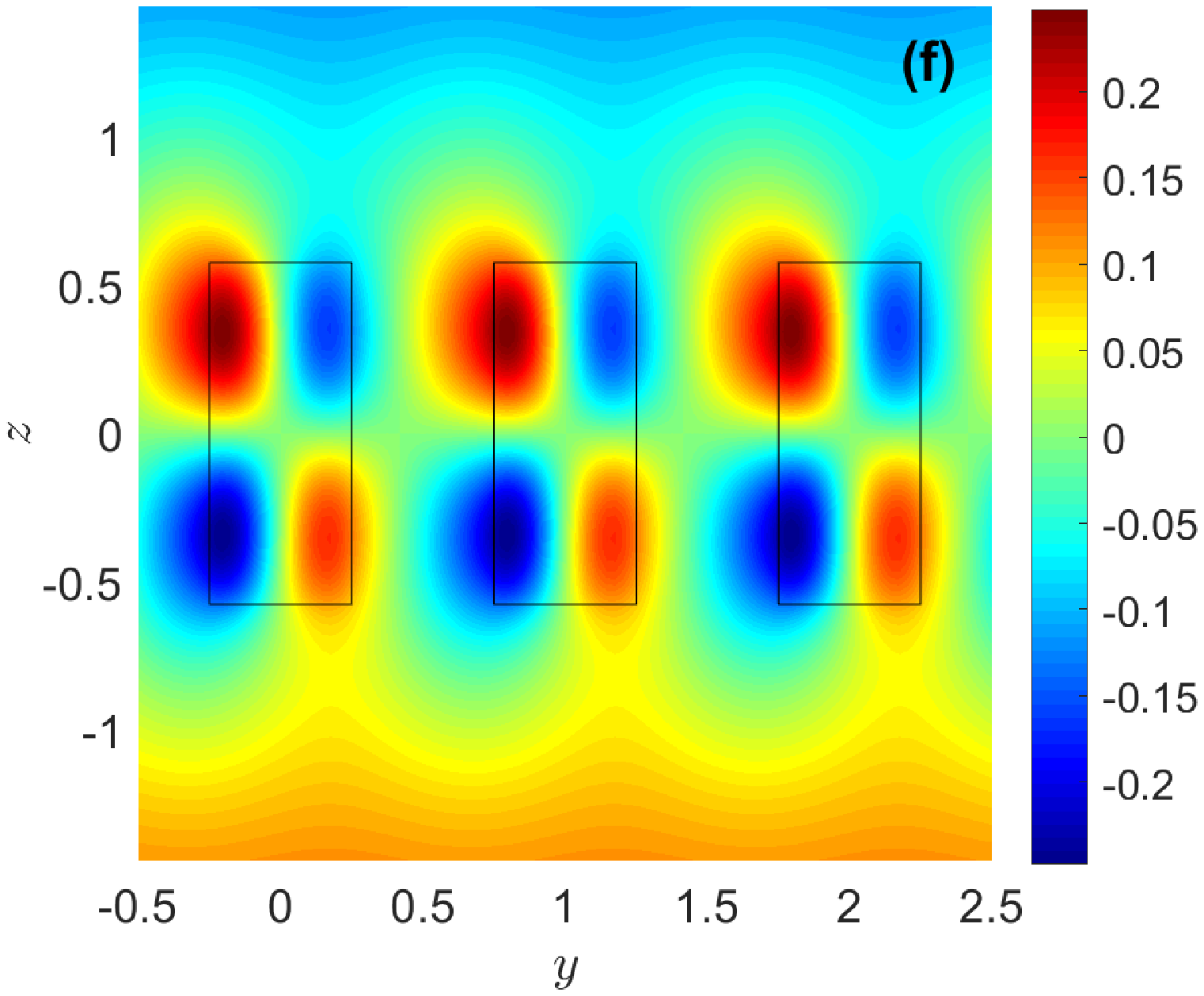}}
	\caption{(a) Dispersion curves around an EP with
          $\beta_*=0.187005\,(2\pi/d)$. (b) The eigenfunction of the
          EP solution. (c), (d), (e), (f) The eigenfunctions at
          points $A$, $B$, $B'$, $A'$ shown in (a),
          respectively.}	\label{Fig2} 
\end{figure*}

More features of the EPs can be understood by closely examining the 
behavior of the eigenmodes in the vicinity of $\beta_*$. In 
Fig.~\ref{Fig2}(a), we show the real and imaginary parts of $k$ 
for two resonant modes around the EP. 
In Figs.~\ref{Fig2}(c)~and~\ref{Fig2}(d),  we show the two
eigenfunctions, $\phi_+$ and $\phi_-$, corresponding to points $A$
and $B$ in Fig.~\ref{Fig2}(a), respectively. As $\beta \to
\beta_*$, those two eigenfunctions coalesce into a single EP
eigenfunction $\phi_*$ as shown in
Fig.~\ref{Fig2}(b). Notice that the field pattern of $\phi_*$
combines the main features of $\phi_-$ and $\phi_+$. For $\beta>
\beta_*$, two different eigenfunctions are recovered, and they are 
shown in Figs.~\ref{Fig2}(e) and \ref{Fig2}(f), corresponding to
points $B'$ and $A'$ in Fig.~\ref{Fig2}(a), respectively. Notice that  
the field patterns for points $A$ and $A'$ are similar, and those for $B$ and
$B'$ are also similar, but their ordering with respect to
$\RE(k)$ is reversed. This switching behavior of the field
patterns as $\beta$ passes through  $\beta_*$ is consistent with
Eq.~\eqref{EqSW}.

To develop a better understanding about  EPs on the periodic slab, we will 
analyze their dependence on parameter $a$ in Sec.~\ref{S5}. 
It will be shown that these EPs continue to exist as $a\to d$, and their
$\beta_*$ and $k_*$ approach the 
light line. It also appears that the limiting points on the light line
are related to some artificial degeneracies of the uniform slab when
it is regarded as a periodic structure. In the next section, we study
these artificial degeneracies and calculate the limiting points on the light 
line. 

\section{Uniform slab\label{S4}}

Referring to Fig.~\ref{Fig1}, by setting $a = d$, we have a uniform
slab with  $\vare = \vare_1$. A guided mode is now given by  
\begin{equation}\label{Eq51}
u(y,z)= \tilde{\phi}(z) e^{i \tilde{\beta} y},
\end{equation}
where $\tilde{\beta}$ is the propagation constant, and
$\tilde{\phi}$ satisfies 
\begin{equation}\label{Eq52}
\frac{d^2\tilde{\phi}}{dz^2}+k^2\vare(z) \tilde{\phi} =
\tilde{\beta}^2 \tilde{\phi}, 
\end{equation}
where $\vare(z)=\vare_1$ for $|z| < h/2$ and $\vare(z)=1$ for $|z|>
h/2$. In addition, $\tilde{\phi}$ must decay to zero as $\abs{z}\to
\INF$. As in previous sections, we only consider
the modes that are odd in $z$, and denote their dispersion relations
as $k = k_m(\tilde{\beta})$ for positive integers $m$. It is
straightforward to show that these dispersion curves touch the light
line when 
\begin{equation}\label{Eq53}
\tilde{\beta}  = k_m(\tilde{\beta})= \tilde{\beta}_m \triangleq \frac{2\pi(m-1/2)}{h\sqrt{\vare_1-1}}.
\end{equation}

\begin{figure}[htb]
 	\centering
 	\includegraphics[width=0.8\linewidth]{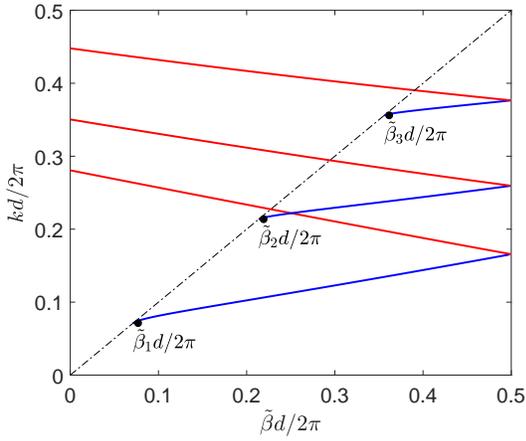}
 	\caption{Folded band structure of a uniform slab with $h= 1.85\,d$. }\label{Fig8}
 \end{figure}

The uniform slab can be regarded as a periodic structure with a
\emph{fictitious} period $d$, then the guided modes can be
written as $u(y,z) =  \phi(y,z) \exp( i\beta y)$, where 
\[
\phi(y,z)=\tilde{\phi}(z) e^{i 2\pi\, l\, y/d },\quad \beta =
\tilde{\beta}- 2\pi l/d, 
\]
 for some integer $l$ such that $\beta\in [-\pi/d,\pi/d]$.
If $|\tilde{\beta}| \leq \pi/d$,  then $l = 0$ and $\beta =
\tilde{\beta}$. Otherwise, $l$ is nonzero and the dispersion curves of
the fictitious periodic slab can be obtained by folding the dispersion
curves of the uniform slab  into the first Brillouin zone. 
This is shown in Fig.~\ref{Fig8}
 for $h = 1.85\,d$, where the solid blue and red curves correspond to
 $l=0$ and $l=1$, respectively. More precisely, the sold red curves
 are 
\[
k = k_m^{({\rm f})}(\beta) \triangleq k_m(2\pi/d-\beta).
\]  

\begin{figure}[htbp]
	\centering
	\includegraphics[width=0.75\linewidth]{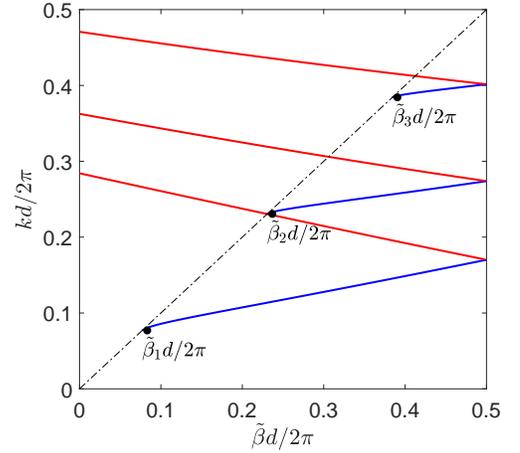}
	\caption{Folded band structure of a uniform slab with $h= 1.7137192\,d$.}\label{Fig9}
\end{figure}

For the case shown in Fig.~\ref{Fig8}, the dispersion curves have no
intersections on the light line. However, it is possible to have 
intersections on the light line if we tune the value of $h$. In
general, we have the following ``limiting degeneracy problem'':
\begin{center}
  \textit{Given integers $m > n > 0$, find $h_*$ such that}
 \begin{equation}\label{Eq54}
     k_n(2\pi/d-\tilde{\beta}_m) =\tilde{\beta}_m. 
 \end{equation}
\end{center}
 For $m=2$ and $n = 1$, we solve the above problem and 
obtain $h_*=1.7137192\,d$. In Fig.~\ref{Fig9}, 
the dispersion curves
are shown for this $h_*$, and the intersection on the light line is 
$k_* = \beta_* =0.2304989\,(2\pi/d)$. 
The results for different pairs $(m,n)$ are listed in Table~\ref{Tab1}. 
\begin{table}[htb]
	\caption{\label{Tab1}Solutions of the ``limiting degeneracy
          problem''.} 
	\begin{ruledtabular}
		\begin{tabular}{ccc}
			$(m,n)$&$h_*/d$ &$k_*d/2\pi$\\\hline
			$(2,1)$&1.7137192 & 0.2304989 \\
			$(3,1)$&3.0874410 & 0.2132351 \\
			$(3,2)$&2.5587359 & 0.2572953 \\
			$(4,1)$&4.4232537 & 0.2083741\\
			$(4,2)$&4.0470027 & 0.2277466\\
			$(4,3)$&3.3127043 & 0.2782292 \\
			$(5,1)$&5.7438584 & 0.2063128 \\
			$(5,2)$&5.4487877 & 0.2174853\\
			$(5,3)$&4.9112722 & 0.2412881 \\
			$(5,4)$&4.0155953 & 0.2951073  \\
		\end{tabular}
	\end{ruledtabular}
\end{table}

\section{Continuation of EPs\label{S5}}

\begin{figure}[htb]
 	\centering
 	\includegraphics[width=0.8\linewidth]{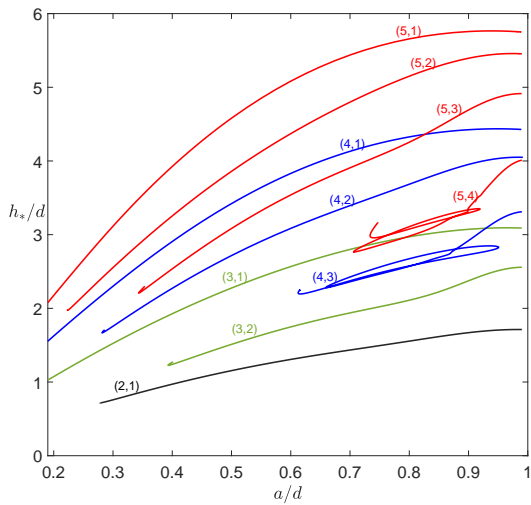}
 	\caption{Slab thickness $h_*$ versus width $a$ for EPs
          satisfying $\beta_* < \mbox{Re}(k_*) < 2\pi/d-\beta_*$.}\label{Fig10} 
 \end{figure}

\begin{figure}[htbp]
	\centering 
	\includegraphics[width=0.8\linewidth]{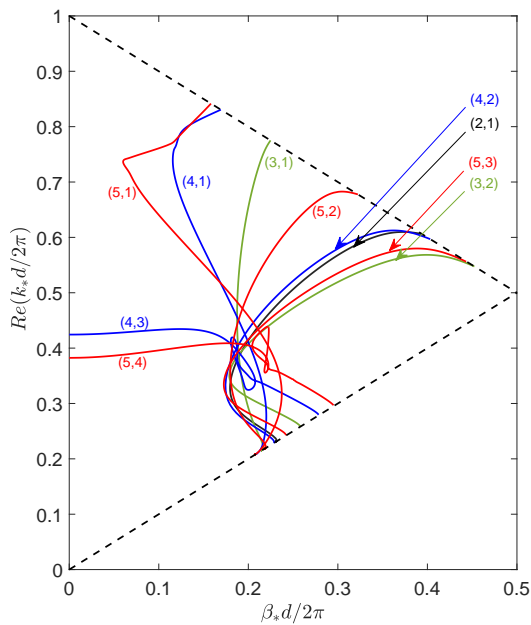}
	\caption{Eigenvalue $k_*$ versus wavenumber $\beta_*$ for the
          same EPs as shown in Fig.~\ref{Fig10}. The curves tend to the light
          line as $a \to  d$.}\label{Fig11}
\end{figure}

\begin{figure}[htbp]
	\centering 
	\includegraphics[width=0.8\linewidth]{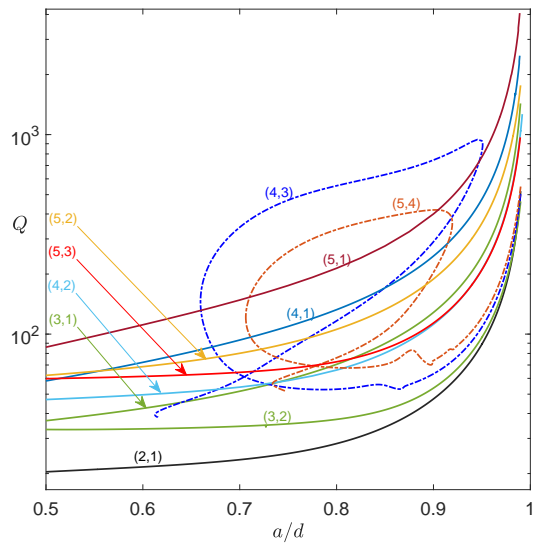}
	\caption{ Quality factors versus $a$ for the same EPs as shown in
          Fig.~\ref{Fig10}. The quality factors diverge as $a\to d$.}\label{Fig12}
\end{figure}

In Sec.~\ref{S3}, we presented one EP for a periodic slab with
$a=0.5d$. In fact,  many different
EPs can be found for the same $a$, but  
typically for different slab thickness. 
To get a more complete picture about the EPs on the periodic slab, 
we use the numerical method of Sec.~\ref{S3} to follow
the EPs in the parameter space of $a$ and $h$, while keeping 
$\vare_1=15.42$ and $\vare_2=1$ fixed. For simplicity, only
those resonant modes that are odd in $z$ are considered. The results 
are shown in Figs.~\ref{Fig10}, \ref{Fig11} and \ref{Fig12}
for $h_*$ vs. $a$, $\mbox{Re}(k_*)$ vs. $\beta_*$, and quality factor
vs. $a$, respectively. 

In Fig.~\ref{Fig10}, we show ten curves in the $a\textendash h$ plane,
where each curve represents one family of EPs  that depend
continuously on $a$ and $h$. These curves are labeled by pairs of integers $(m,n)$. Notice
that as  $a\to d$, all these curves tend to the values of $h_*$ listed
in Table~\ref{Tab1}, according to their corresponding labels. The
$\beta_*$ and $k_*$ values of the ten families of EPs are shown in
Fig.~\ref{Fig11}. As $a\to d$, the curves in Fig.~\ref{Fig11}
approach the light line exactly at the values of $k_*$ listed in
Table~\ref{Tab1}.  Therefore, it can be argued that all these EPs
{\em originate} from the artificial degeneracies of the guided modes on
the uniform slab. In Fig.~\ref{Fig10}, we observe that
different curves, e.g., $(4,1)$ and $(5,3)$, may intersect. This
simply means that for the same periodic slab, there are two EPs with
different $\beta_*$ and $k_*$. Notice that
the curves $(4,3)$ and $(5,4)$ can intersect with themselves. 
On a periodic slab corresponding to such an intersection, there are
two EPs that belong to the same family, but their $\beta_*$ and $k_*$
are again different.  
The left end points of the 
curves in Fig.~\ref{Fig10} correspond to either $\beta_* = 0$, or 
$\beta_* = 2\pi/d - \mbox{Re}(k_*)$ which is the opening line of the 
second radiation channel. EPs exist beyond this line, but they
tend to have lower quality factors. 
The quality factors of the ten families of EPs are shown in 
Fig.~\ref{Fig12}. It can be observed that the quality factors of all 
EPs diverge as $a\to d$, and they are relatively small for smaller 
values of $a$.  

\begin{figure*}[htb]
	\subfloat{\includegraphics[width=0.33\textwidth]{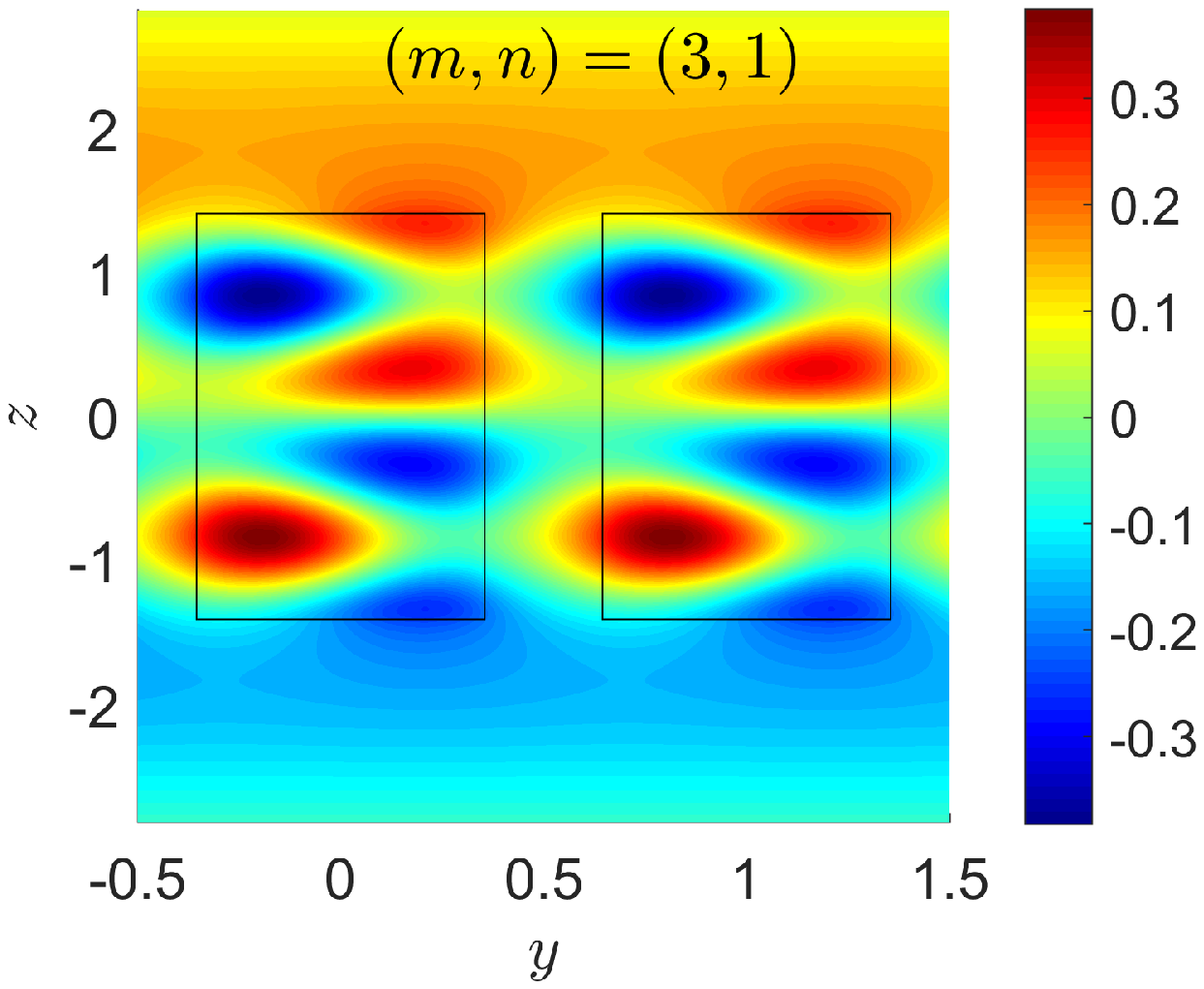}}\hfill
	\subfloat{\includegraphics[width=0.33\textwidth]{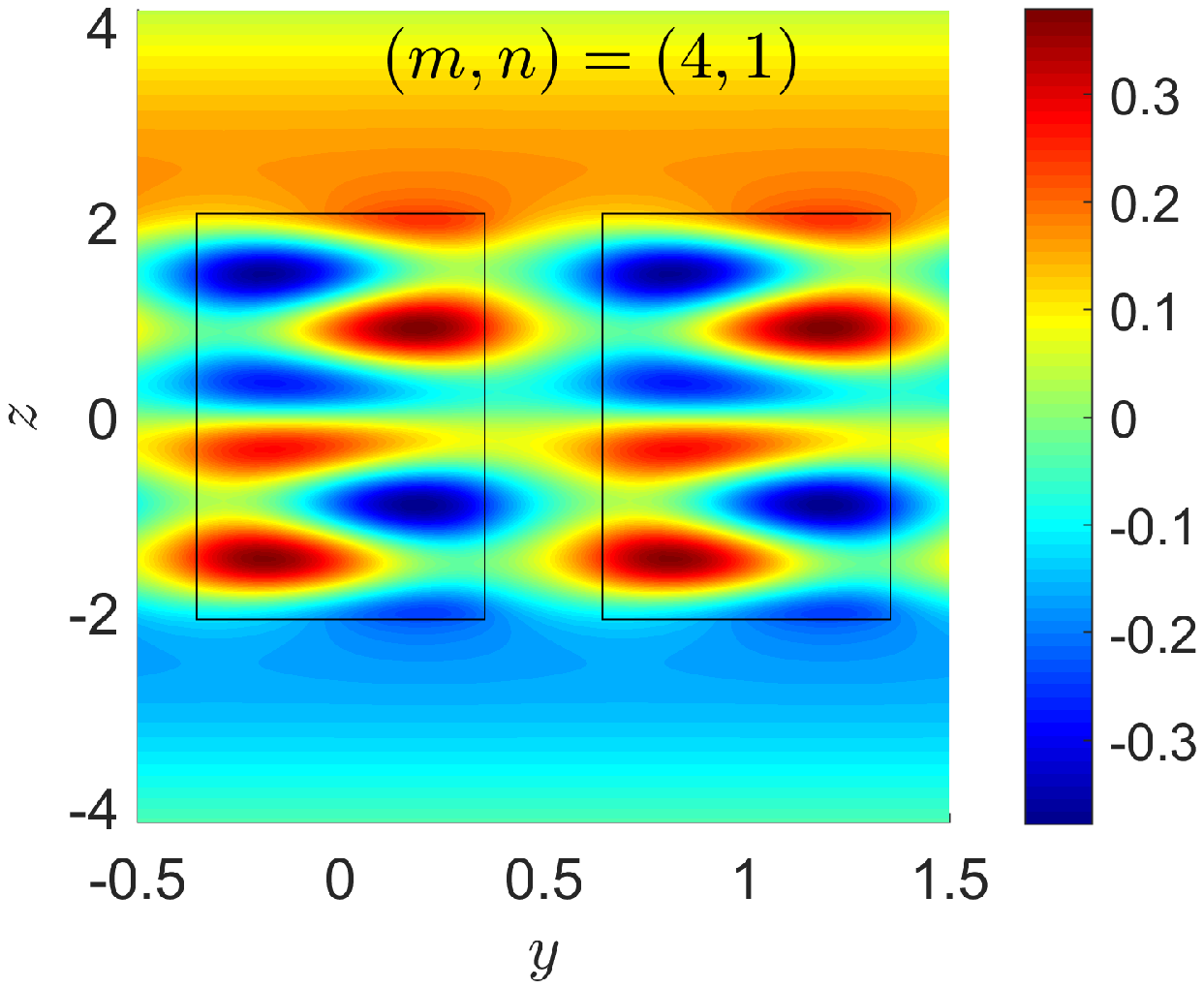}}\hfill
	\subfloat{\includegraphics[width=0.33\textwidth]{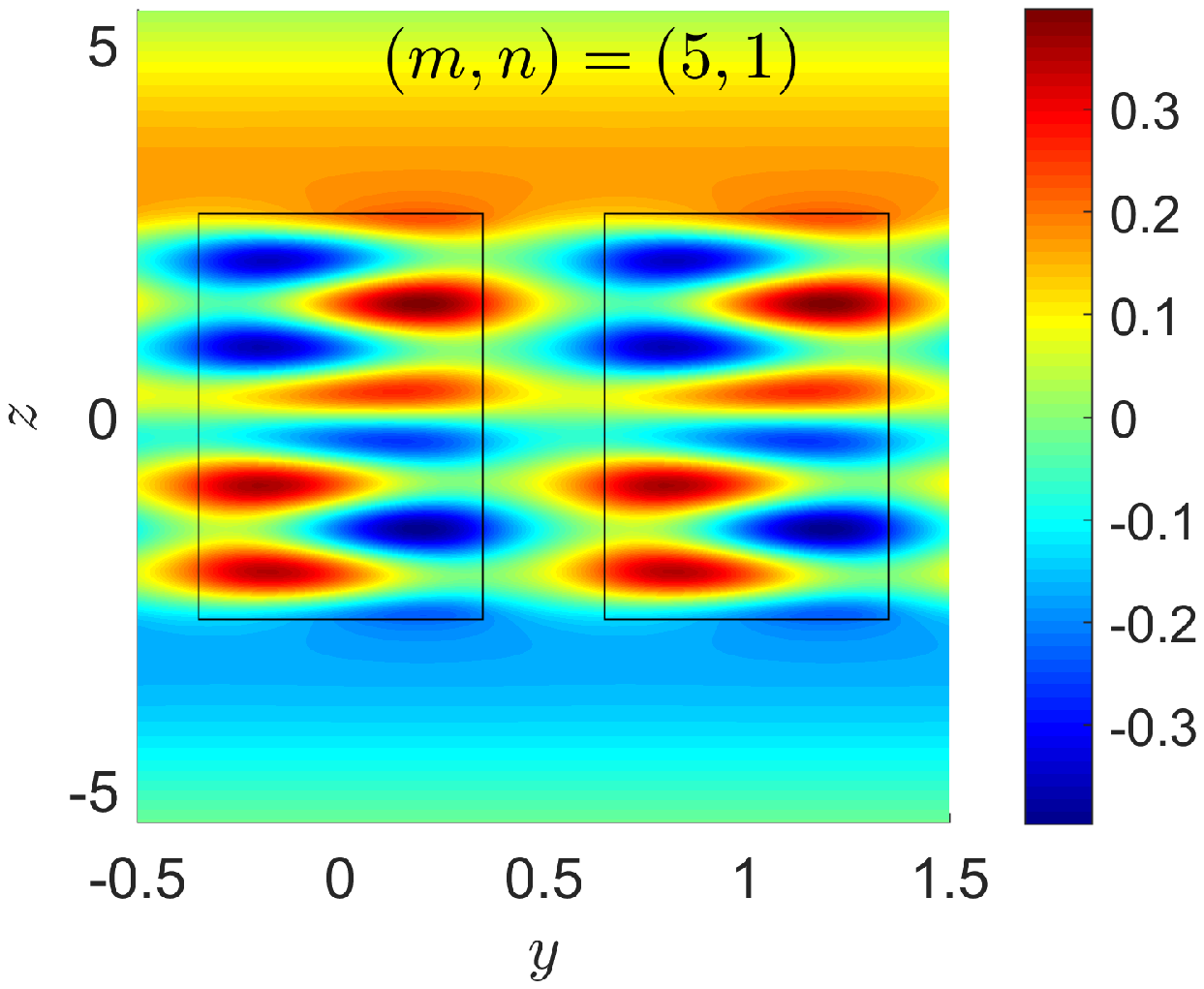}}\hfill\\[-2ex]
	
	\subfloat{\includegraphics[width=0.33\textwidth]{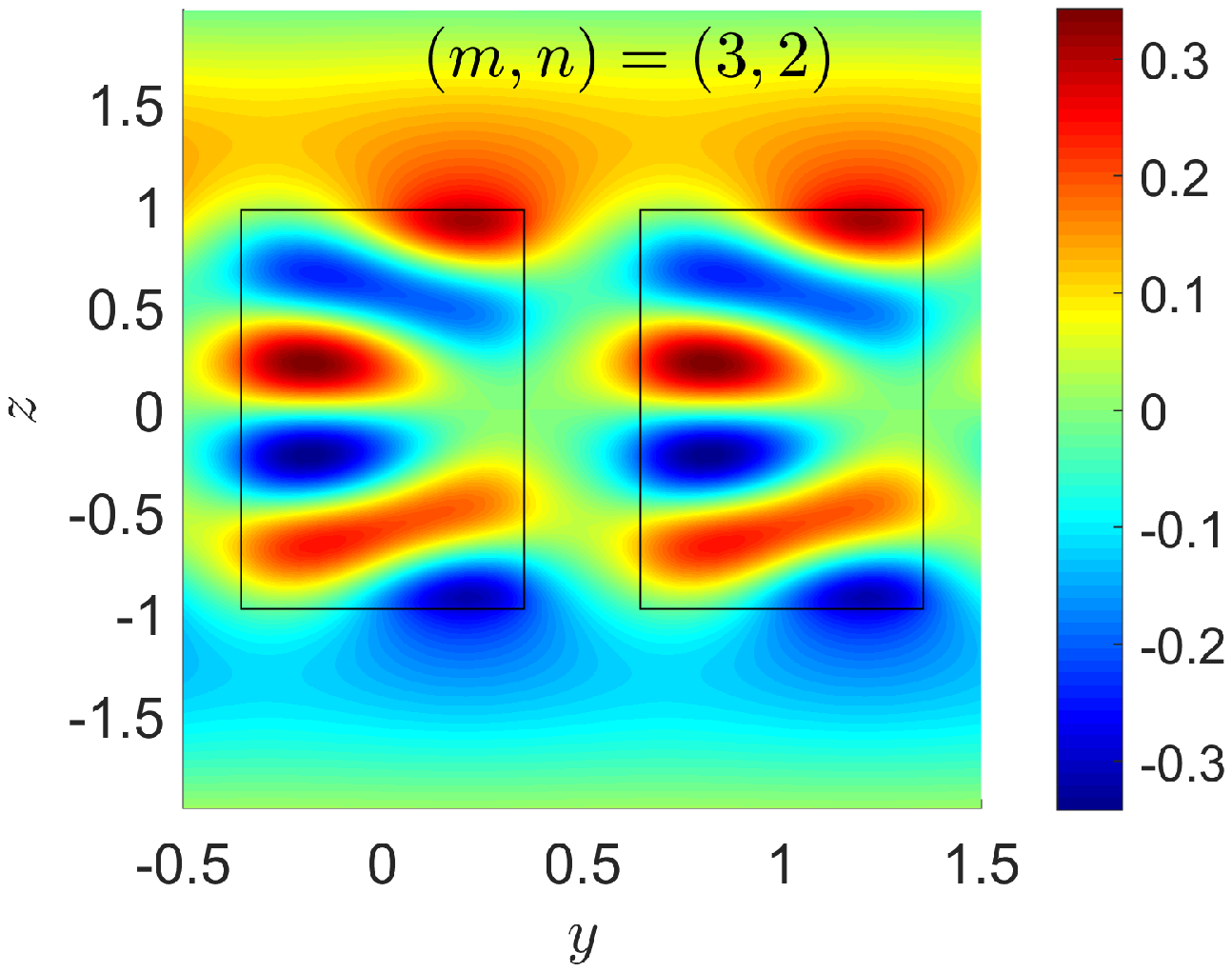}}\hfill
	\subfloat{\includegraphics[width=0.33\textwidth]{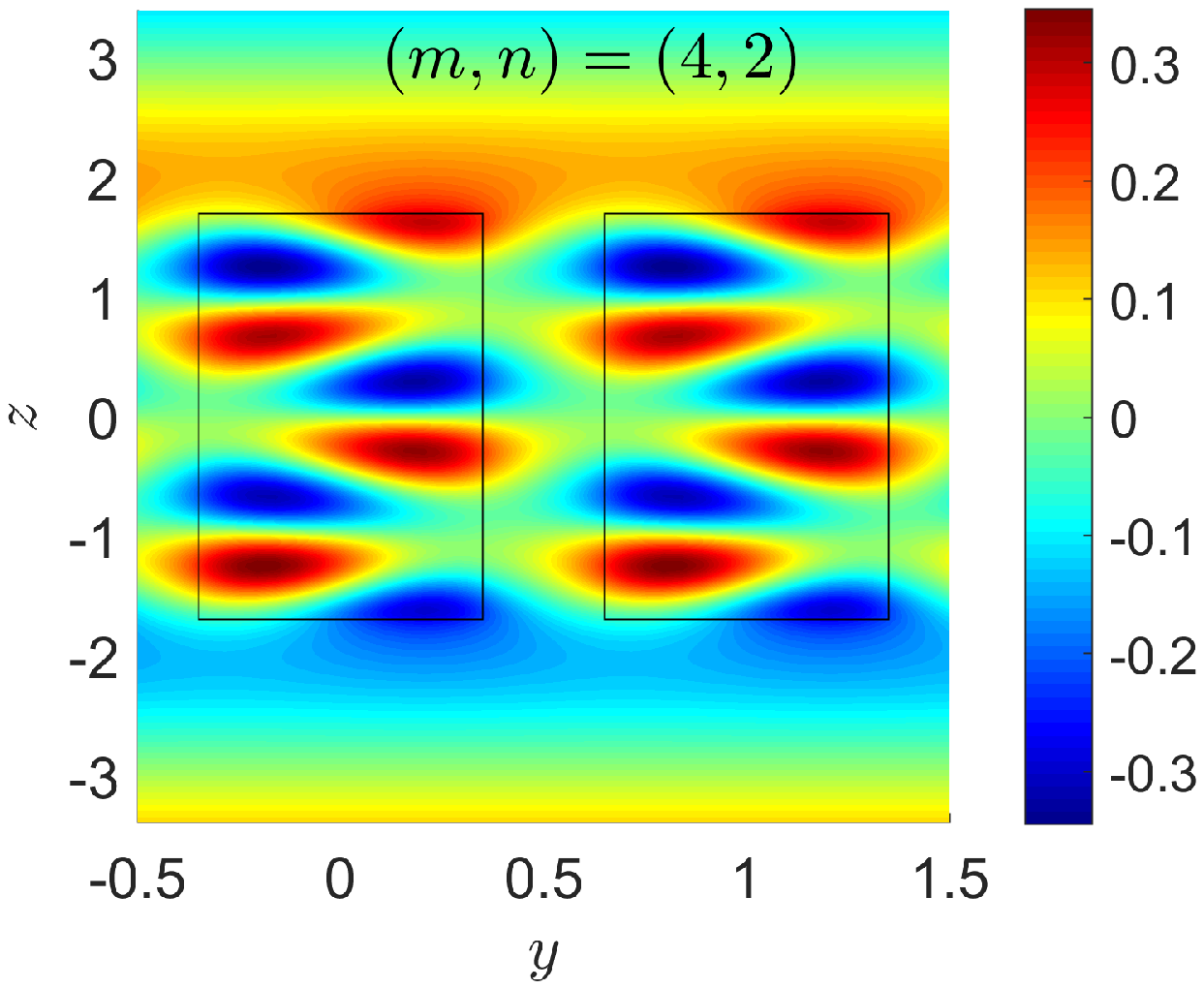}}\hfill
	\subfloat{\includegraphics[width=0.33\textwidth]{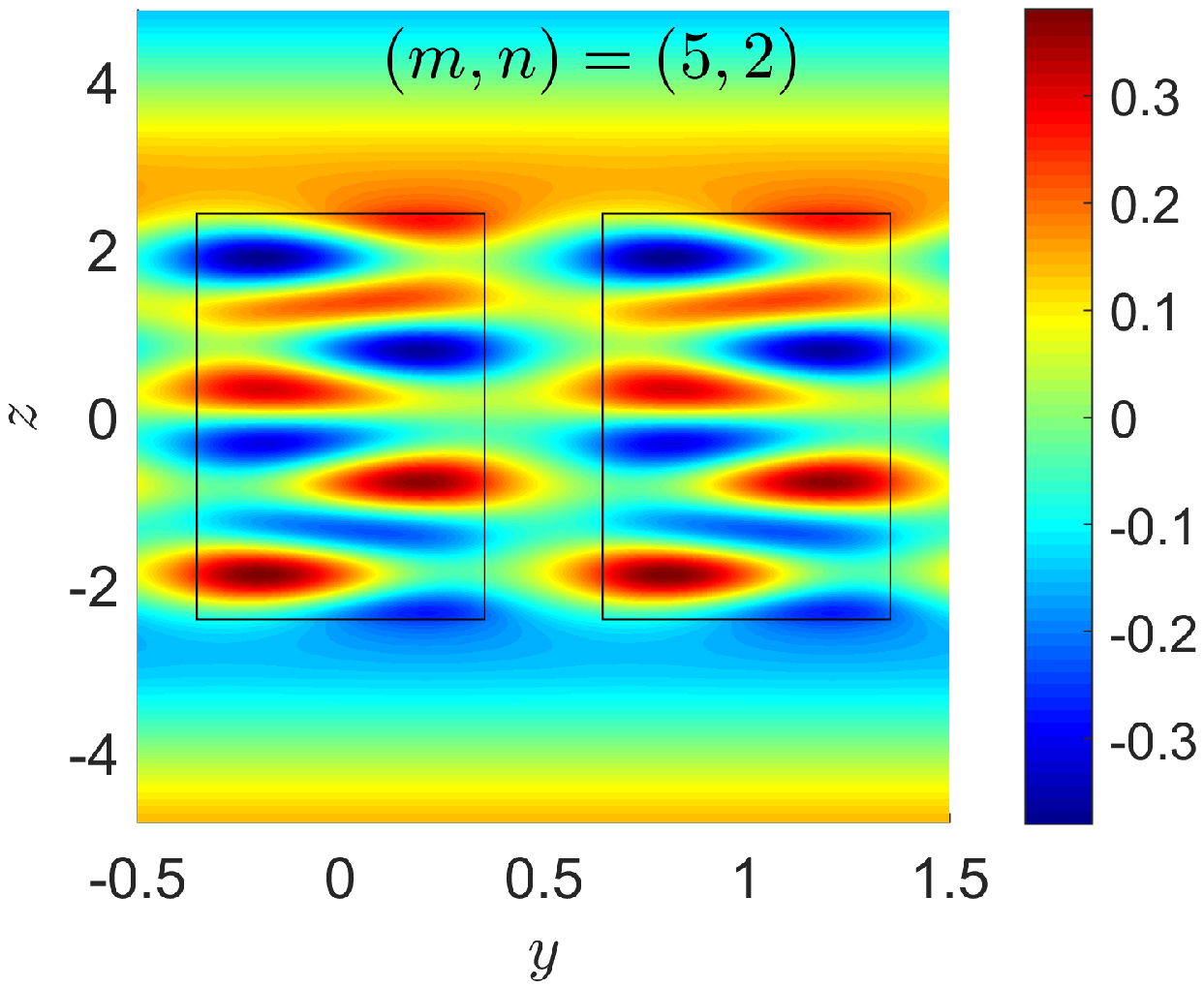}}\hfill\\[-2ex]
	
	\subfloat{\includegraphics[width=0.33\textwidth]{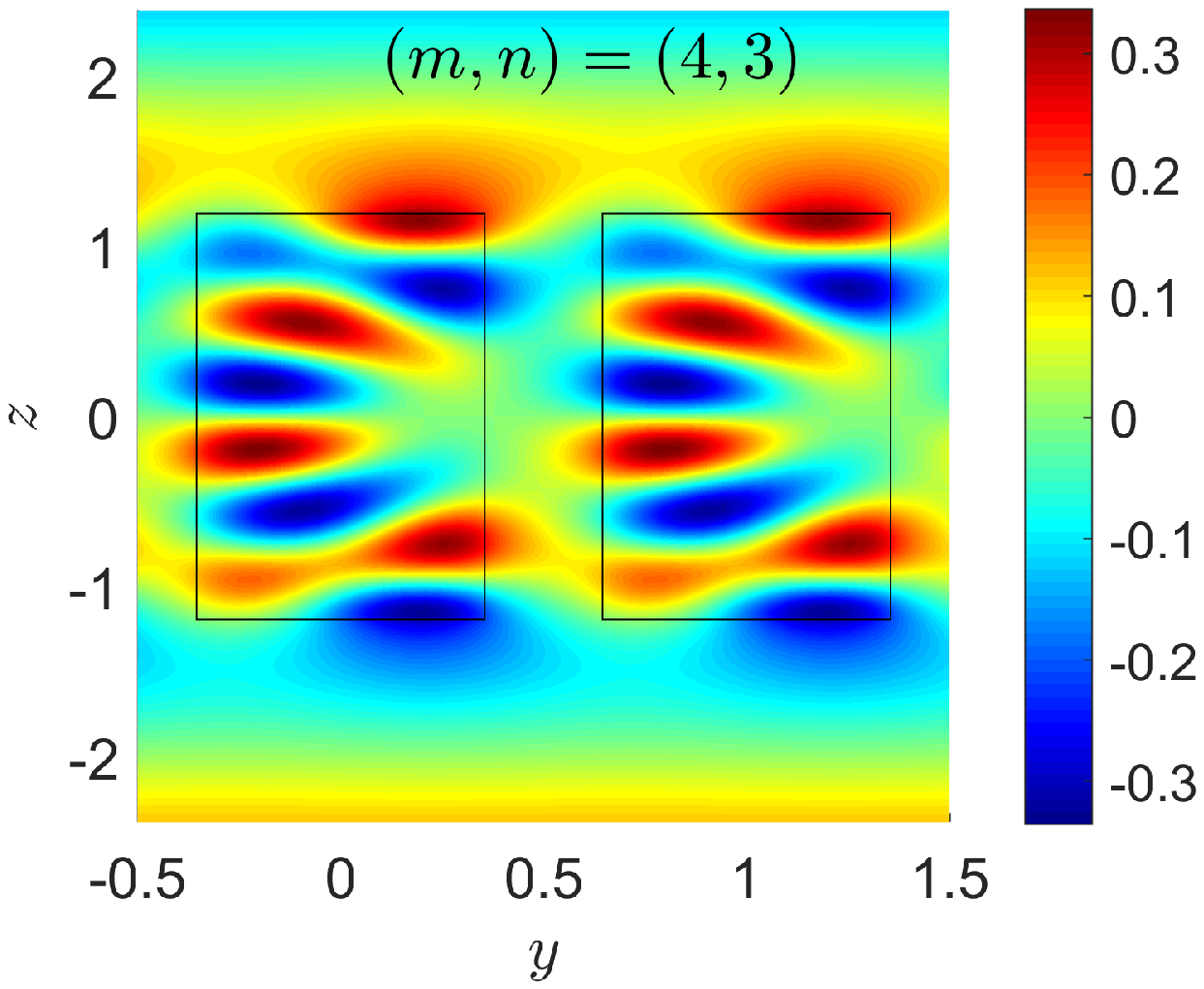}}\hfill
	\subfloat{\includegraphics[width=0.33\textwidth]{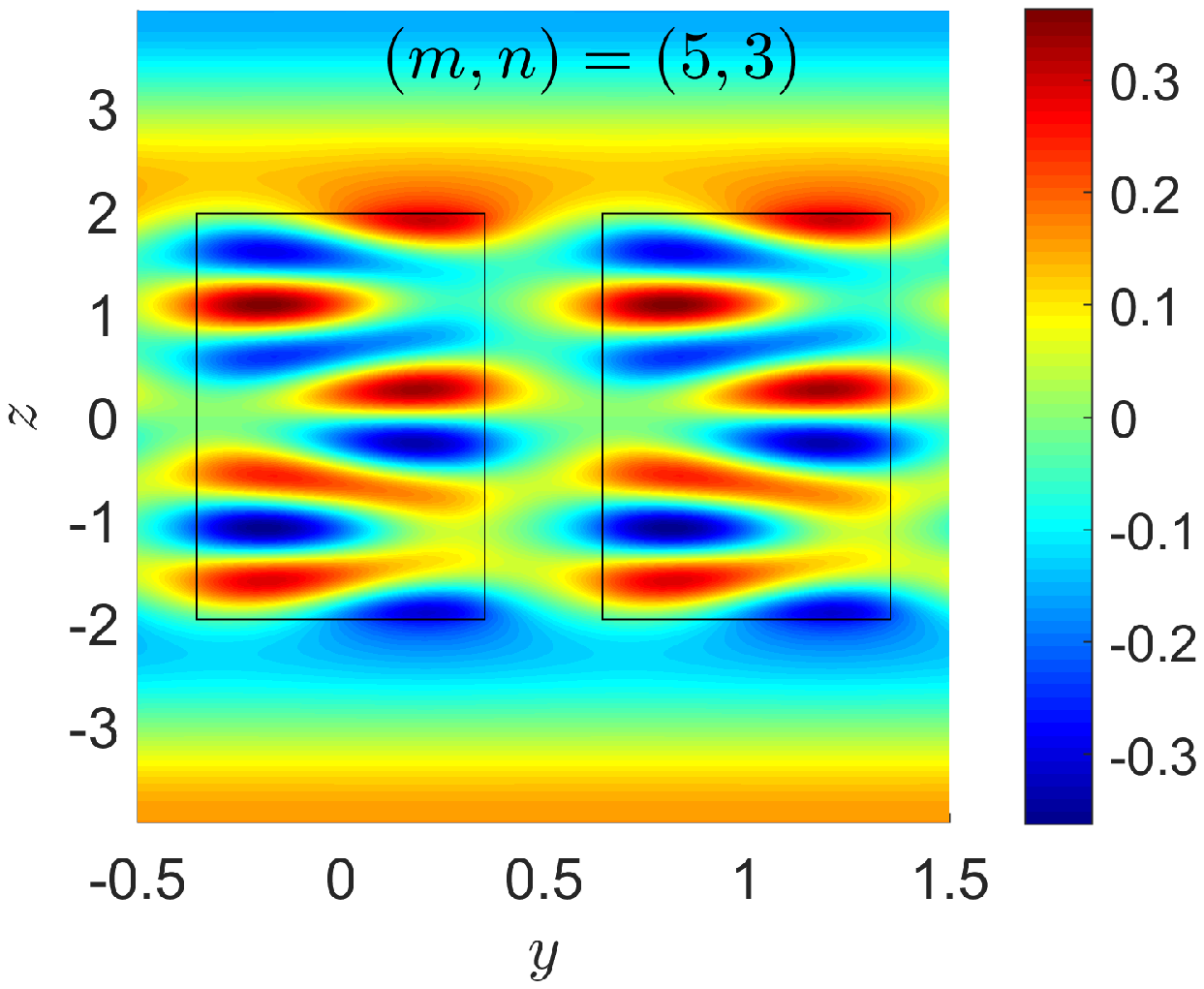}}\hfill
	\subfloat{\includegraphics[width=0.33\textwidth]{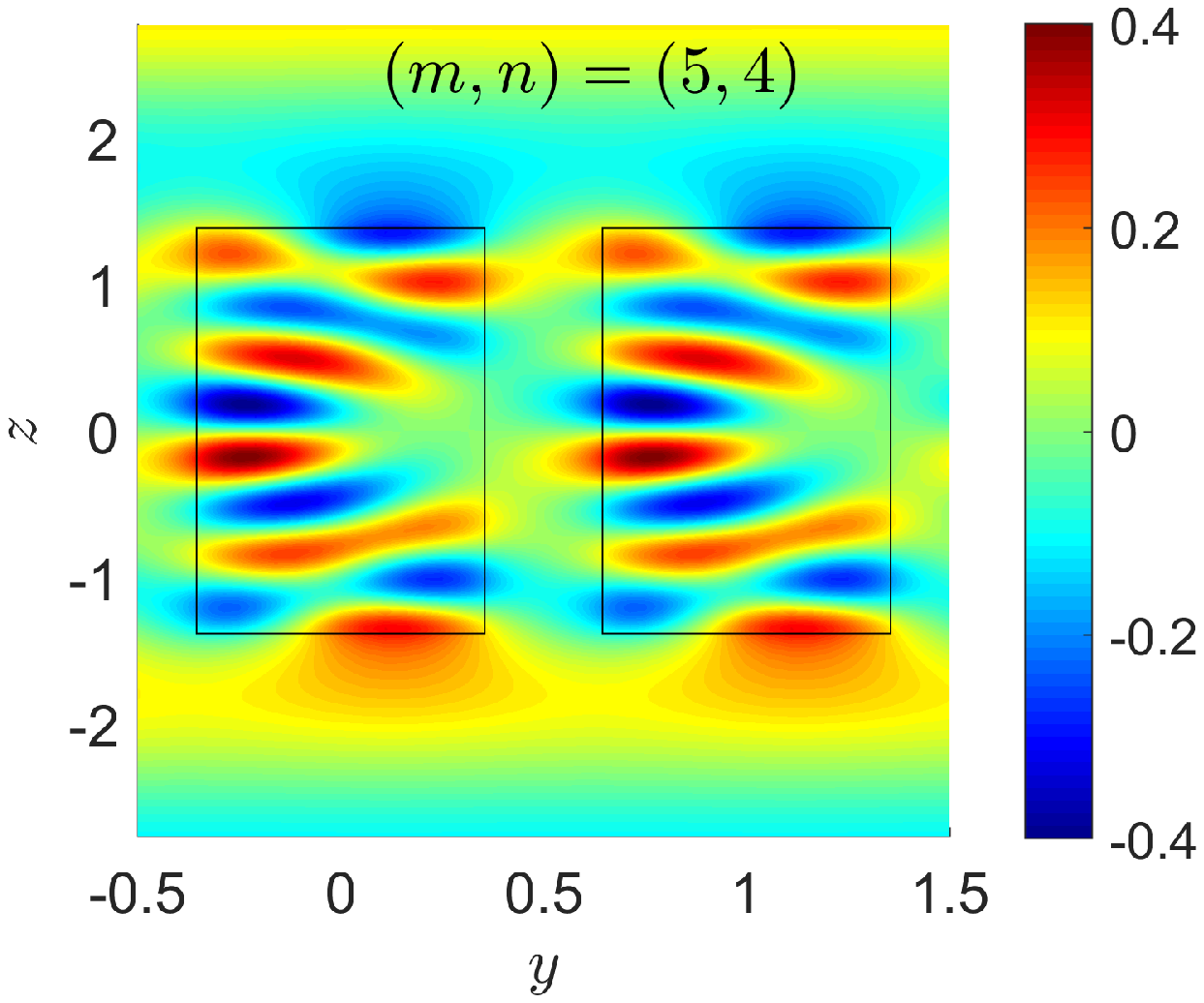}}\hfill
	
	\caption{Eigenfunctions $\phi_*$ for EPs on periodic slabs with
          $a = 0.7d$ and different $h_*$. The EPs are labeled by integer pairs $(m,n)$ as
          in Fig.~\ref{Fig10}.}	\label{Fig13}  
\end{figure*}

In Fig.~\ref{Fig13}, 
we show the field patterns of nine EPs on periodic slabs with $a =
0.7d$. The case for $(m,n) = (2,1)$ is not shown, since it is similar
to the one in Fig.~\ref{Fig2}(b) for $a = 0.5d$.  Notice that different scales in $z$ are used in the different panels, 
since the values of $h_*$ are different for different EPs. 
It should be pointed out that the wave field pattern varies
continuously on each curve shown in Fig.~\ref{Fig10}, and the main features,
such as the number of polarity changes along the $y$ and  $z$ axes,
are preserved when $(a,h_*)$ moves along each curve. 

\begin{figure}[htb]
  	\subfloat{\includegraphics[width=0.4\textwidth]{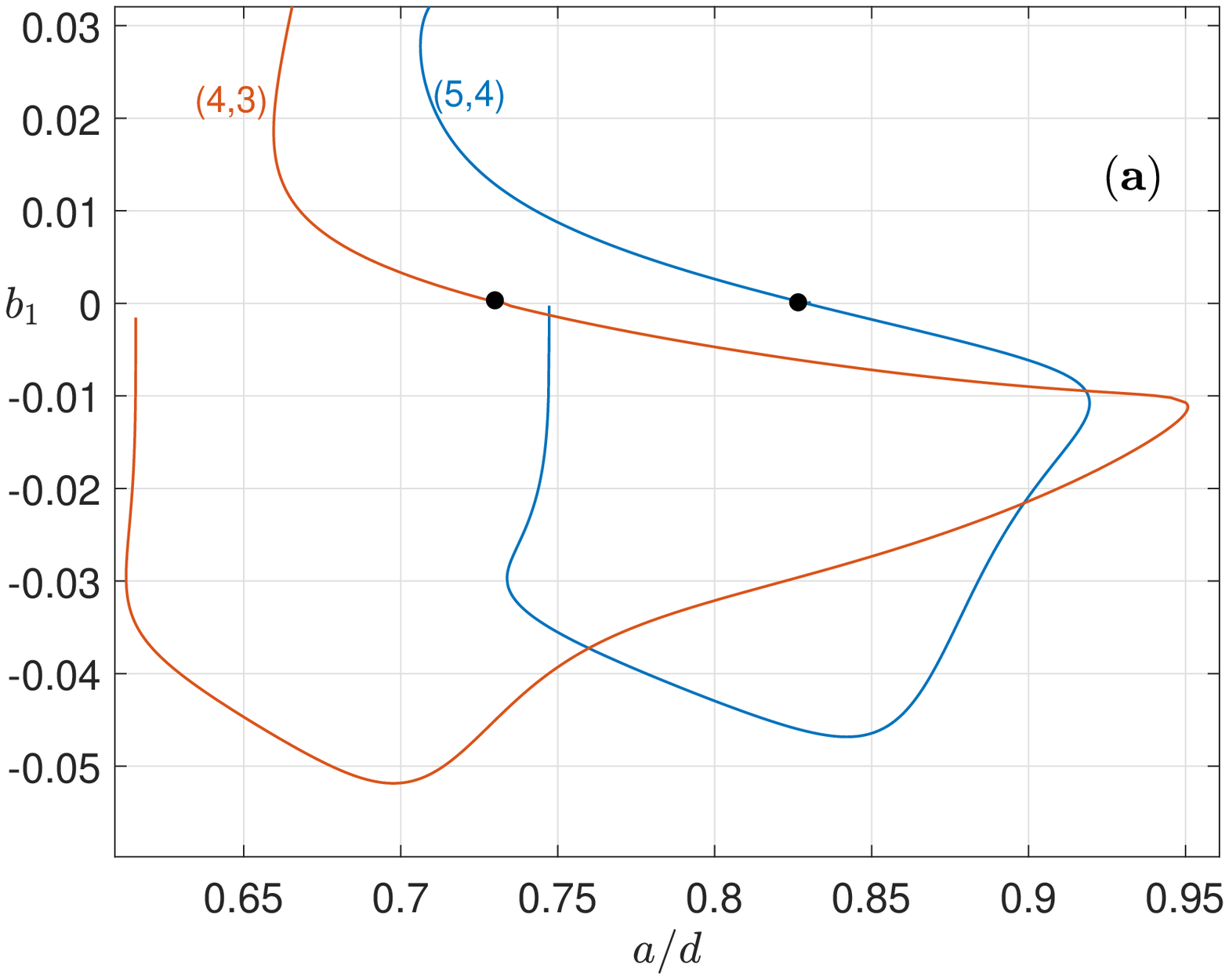}}\hfill 
  	\subfloat{\includegraphics[width=0.4\textwidth]{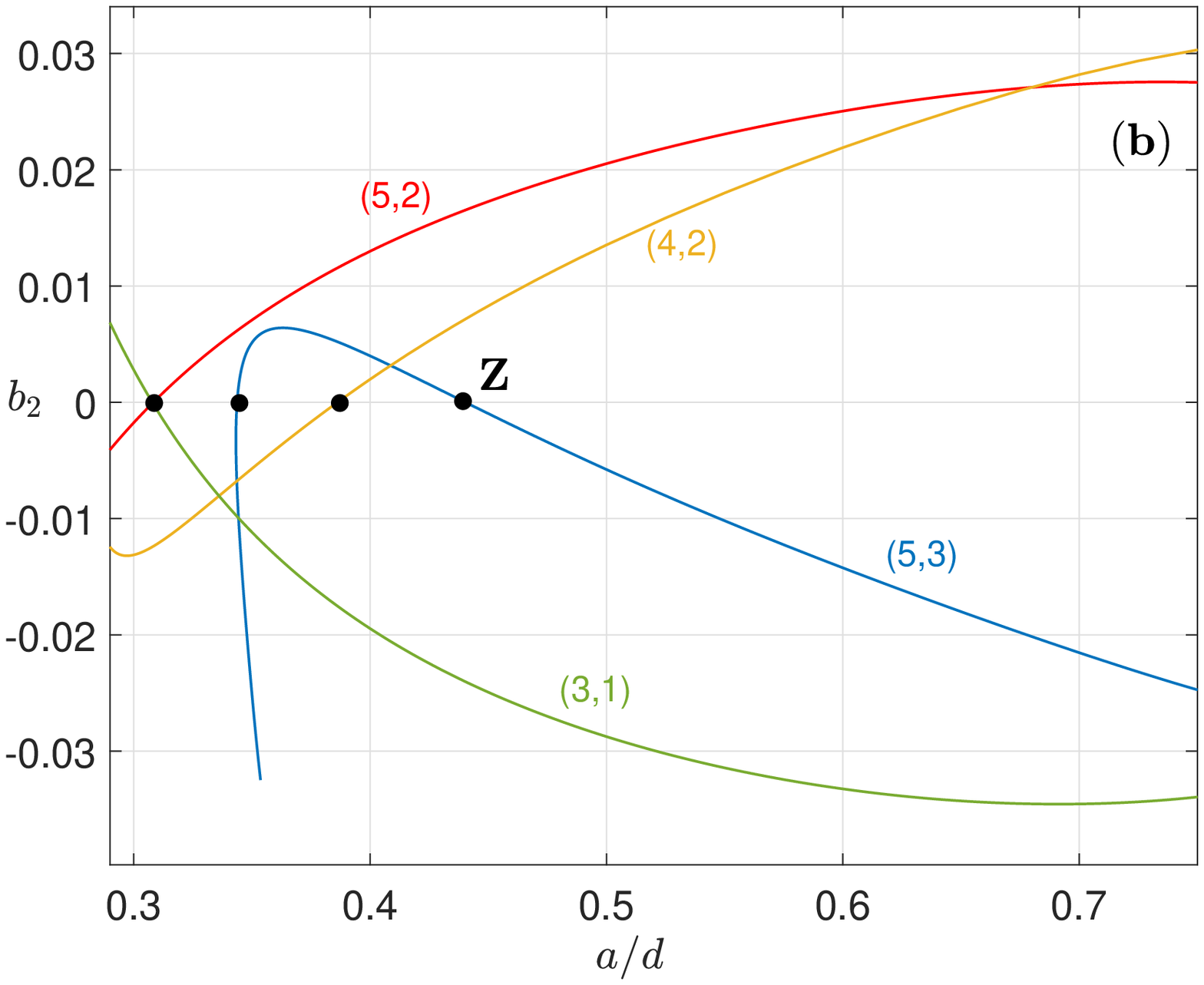}}
  	\caption{(a) Coefficient $b_1$ versus $a$. (b) Coefficient $b_2$
          versus $a$. Solid black dots are zeros of $b_1$ or $b_2$.}
	\label{Fig14}
  \end{figure}

\begin{figure}[htb]
    \subfloat{\includegraphics[width=0.45\textwidth]{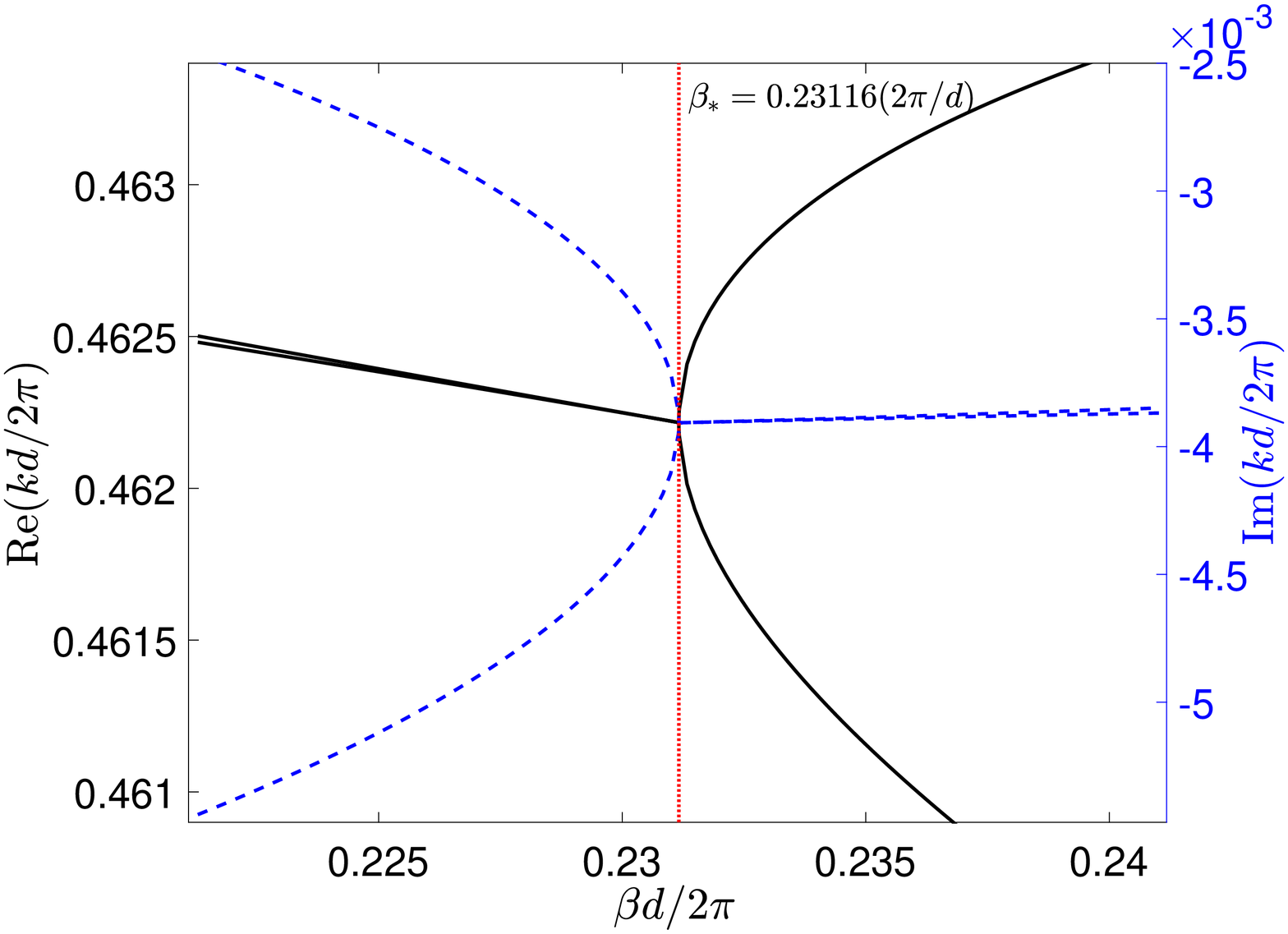}}
	\caption{One-sided square-root splitting of $\RE(k)$ (solid black
          curves) and $\IM(k)$ (dashed blue curves) for an EP with
          $b_2 = 0$, i.e., point $Z$ in Fig.~\ref{Fig14}\,(b).}	\label{Fig15}  
\end{figure}

Near a second-order EP, the dispersion curves exhibit a square-root
splitting. One example is shown in Fig.~\ref{Fig2}(a), where a 
square-root splitting can be observed for both real and imaginary
parts of $k$, and for both $\beta > \beta_*$ and $\beta <
\beta_*$. For this particular case, $\mbox{Re}(k)$ has a stronger
splitting for $\beta > \beta_*$, and $\mbox{Im}(k)$ has a stronger
splitting for $\beta < \beta_*$, since $|b_1| =|c_2| > |b_2| = |c_1|$.
For some special EPs, $b_1$ or $b_2$ can be exactly zero. 
In  Figs.~\ref{Fig14}\,(a) and \ref{Fig14}(b), 
we show $b_1$ and $b_2$ vs. $a$ for a few families of EPs. It is clear 
that $b_1$ or $b_2$ can be zero for some special values of $a$. 
If $b_1$ or $b_2$ is zero, the square-root splitting becomes
one-sided. For example, if $b_2=c_1=0$, then $\RE(k)$ has a
square-root splitting only for $\beta>\beta_*$, and  $\IM(k)$ has a
square-root splitting only for $\beta<\beta_*$. A particular example is
shown in Fig.~\ref{Fig15}, 
and it corresponds to point $Z$ marked in Fig.~\ref{Fig14}(b). 
Actually, even though $b_2=0$, $\IM(k)$ still shows a splitting for
$\beta>\beta_*$, but it is a weaker linear splitting proportional to
$\beta-\beta_*$. The same is true for $\RE(k)$ and $\beta < \beta_*$. 

\section{Conclusion\label{S6}}

Electromagnetic resonant modes on open dielectric structures are
solutions of a non-Hermitian eigenvalue problem derived from the
Maxwell's equations. EPs of resonant modes are special degenerate
states for which both the eigenvalues and the eigenfunctions coalesce.
In optical systems, EPs have given rise to many interesting wave
phenomena and some important applications. In this paper, we
investigated exceptional points on a simple dielectric periodic slab.
An efficient numerical method for computing second-order EPs was
developed and used to calculate families of EPs that vary continuously
with structural parameters. Due to the extra degree of freedom related
to the Bloch wavenumber, it is possible to find EPs by tuning only one
parameter of the periodic structure.  It is worth mentioning that
analytic results have been obtained for the limit $a\to d$, i.e., the
periodic slab approaching a uniform one.  It was shown that the EPs
tend to some points on the light line in this limit, and these points
are related to some artificial degeneracies of the band structure of a
uniform slab when it is regarded as a periodic one. 

Our results show that the EPs have rather complicated dependence on
the geometric parameters of the periodic slab. For simplicity, we have
concentrated on the odd resonant modes in the $E$ polarization for a
very simple periodic slab. Further studies are needed to understand EPs
on more general and three-dimensional structures, to reveal their
properties including existence and robustness, to find out their
impact on transmission spectra, reflection spectra and field
enhancement, and to realize more valuable applications. 

\begin{acknowledgments}
The second author acknowledges support from the Research Grants 
Council of Hong Kong Special Administrative Region, China (Grant
No. CityU 11304117).
\end{acknowledgments}

\appendix

\section{Numerical methods}

\begin{figure}[htbp]
	\centering 
	\includegraphics[width=0.90\linewidth]{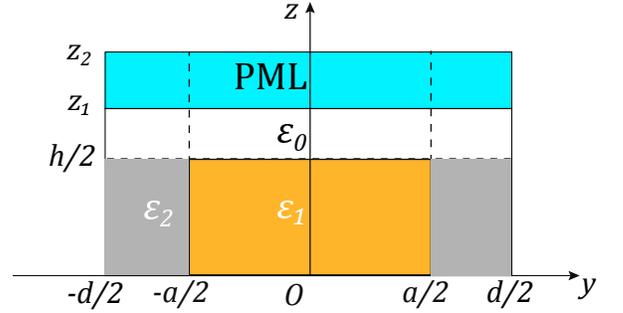}
	\caption{One period of the periodic slab divided into subdomains of
          constant permittivity.} 
	\label{Fig3}
\end{figure}

The eigenvalue problem for resonant modes on the periodic slab can be
solved by a linear scheme based on the Chebyshev pseudospectral
method \cite{Tref} and the perfectly matched layer (PML)  technique
\cite{Berenger,PMLChew}. 
For the periodic function $\phi$ given in Eq.~\eqref{Eq2},
the Helmholtz equation (\ref{Eq1}) becomes 
\begin{equation}\label{Eq7}
\frac{\Par^2\phi}{\Par z^2}+\frac{\Par^2\phi}{\Par y^2}+2i\beta \frac{\Par\phi}{\Par y} -\beta^2\phi  = -k^2\vare \phi.
\end{equation}
If the modes are odd in $z$, it is only necessary to consider
$z>0$ together with a zero boundary condition at $z=0$. 
The $z$ axis is truncated at $z=z_2$ with a PML for $z_1 < z < z_2$ 
as shown in Fig.~\ref{Fig3}. 
The PML replaces $z$ by $\hat{z}$ or 
$dz$ by $d\hat{z} = s(z)\,dz$ for a complex function $s(z)$.
Hence, Eq.~\eqref{Eq7} becomes
\begin{equation}\label{Eq8}
\frac{1}{s(z)}\frac{\Par}{\Par
  z}\left[ \frac{1}{s(z)}\frac{\Par\phi}{\Par
    z}\right] +\frac{\Par^2\phi}{\Par y^2}+2i\beta \frac{\Par\phi}{\Par
  y} -\beta^2\phi  = -k^2\vare \phi. 
\end{equation} 
In addition, $\phi$ must satisfy the boundary conditions
\begin{equation}\label{Eq9}
\phi(y,0)=\phi(y,z_2)=0, 
\end{equation}
and periodic conditions
\begin{eqnarray}
&& \phi(-d/2,z) =\phi(d/2,z), \\
\label{Eq10}
&& \frac{\Par \phi}{\Par y}(-d/2,z) =\frac{\Par \phi}{\Par y}(d/2,z).
\end{eqnarray}

We use the Chebyshev pseudospectral method \cite{Tref} to discretize Eq.~\eqref{Eq8} on each rectangular subdomain shown in
Fig.~\ref{Fig3},  and impose field continuity conditions and
conditions~\eqref{Eq9}-\eqref{Eq10} on the boundaries of 
the subdomains. The result is a linear matrix eigenvalue problem of the form 
\begin{equation}\label{Eq11}
\begin{aligned}
{\bm L}  {\bm \phi} = k^2 {\bm \phi},
\end{aligned}
\end{equation}
where $  {\bm \phi}$ is a vector containing the values of
$\phi$ at the interior Chebyshev collocation points. 


The eigenvalue problem for resonant modes on the periodic slab can also
be solved by a nonlinear scheme based 
on the mode matching method. For the odd (in $z$) modes, the structure
is divided 
into two layers given by $0<z<h/2$ and $ z > h/2$, 
and the dielectric function is $\vare^{(1)}(y)$ and $\vare^{(2)}(y)$
in these two layers, respectively. For given $\beta$ and $k$, we can
expand $\phi$ in each layer in one-dimensional eigenmodes. 
The eigenvalue problem in the $l$-th layer is 
\[
\left[ \frac{d^2}{dy^2}+2i\beta \frac{d}{dy} +k^2\vare
  ^{(l)}(y)-\beta^2\right] \psi^{(l)} = [ \eta^{(l)} ]^2 \psi^{(l)},
\]
subject to the periodic boundary conditions
\begin{eqnarray*}
&& \psi^{(l)}(-d/2) =
  \psi^{(l)}(d/2), \\
&&  \frac{d\psi^{(l)}}{dy} (-d/2) =
  \frac{d\psi^{(1)}}{dy} (d/2).
\end{eqnarray*}
Solving the above by the Chebyshev collocation
method~\cite{Tref}, we obtain 
\[
\set{ \psi^{(l)}_j,\, \eta^{(l)}_j  }_{j=1}^N, \quad l = 1, 2,
\]
for a positive integer $N$ and discretization points $\{ y_p
\}_{p=1}^N$. 

With the above eigenmodes, $\phi$ can be expanded as 
\begin{eqnarray*}
 && \phi(y,z)  = \sum_{j=1}^N 
\frac{c_j \sin(\eta^{(1)}_j z)}{\sin(0.5\eta^{(1)}_j h)}
\psi^{(l)}_j(y), \ \  0\leq z < \frac{h}{2}, \cr
&& \phi(y,z) = \sum_{j=1}^N d_j e^{i\eta^{(2)}_j(z-h/2)}
   \psi^{(2)}_j(y),  \quad  z> \frac{h}{2}.
\end{eqnarray*}
Enforcing the continuity of $\phi$ and $\partial_z \phi$ at $y=y_p$ and
$z=h/2$ for $1\le p \le N$, we obtain the following linear system
\begin{equation}
\label{Eq14}
{\bm A} 
\begin{bmatrix}
\mathbf{c} \cr
\mathbf{d}
\end{bmatrix}
= 
\begin{bmatrix}
{\bm A}_{11} & {\bm A}_{12} \\
{\bm A}_{21} & {\bm A}_{22}
\end{bmatrix} 
\begin{bmatrix}
{\bm c}\cr
{\bm d}
\end{bmatrix}
 = {\bm 0}, 
\end{equation}
where ${\bm c}$ and ${\bm d}$ are column vectors of $c_j$ and 
$d_j$ ($1\le j  \le N$), respectively, ${\bm A}$ is a $ 2\times 2$
block matrix depending on $\beta$ and $k$, and 
the $(p,j)$ entries of the matrix blocks are 
\begin{eqnarray*}
&& {\bm A}_{11}(p,j) = \psi^{(1)}_j(y_p), \cr
&& {\bm A}_{12}(p,j) = -\psi^{(2)}_j(y_p),  \cr
&& {\bm A}_{21}(p,j) =
\eta^{(1)}_j\cot(0.5\eta^{(1)}_j h)\psi^{(1)}_j(y_p), \cr
&& {\bm A}_{22}(p,j) =  -i\eta_j^{(2)} \psi^{(2)}_j(y_p).
\end{eqnarray*}
Equation~\eqref{Eq14} has a nontrivial solution, only when
${\bm A}$ is singular. Therefore, we can find the complex $k$
from the condition $\lambda_1({\bm A}) = 0$, where $\lambda_1(A)$
is the eigenvalue of ${\bm A}$ with the smallest magnitude. 

\bibliography{MyBib}

\end{document}